\documentclass[11pt, a4paper]{article}

\usepackage{color}
\usepackage{graphicx}
\usepackage{amssymb}
\usepackage{amsmath}
\usepackage[english]{babel}
\usepackage[T1]{fontenc}
\usepackage[noadjust]{cite}
\usepackage{hyperref} 

\usepackage[utf8]{inputenc}
\usepackage{authblk}

\newcommand\eeq{\end{equation}}
\newcommand\beq{\begin{equation}}
\newcommand\eea{\end{eqnarray}}
\newcommand\bea{\begin{eqnarray}}

\begin{document}

\linespread{1.1}

\title{ \color{red} \bf Dark matter and dark radiation   \\ from evaporating Kerr primordial black holes}

\author[1]{ {\Large Isabella Masina} \thanks{masina@fe.infn.it}}

\affil[1]{\small Dip. di Fisica e Scienze della Terra, Ferrara University and INFN, Ferrara, Italy }

\date{}

\maketitle

\begin{abstract}
The mechanism of the generation of dark matter and dark radiation from the evaporation of primordial black holes is very interesting.
We consider the case of Kerr black holes to generalize previous results obtained in the Schwarzschild case. 
For dark matter, the results do not change dramatically and the bounds on warm dark matter apply similarly: 
in particular, the Kerr case cannot save the scenario of black hole domination for light dark matter.
For dark radiation, the expectations for $\Delta N_{eff}$ do not change significantly with respect to the Schwarzschild case, 
but for an enhancement in the case of spin 2 particles: in the massless case, however, the projected experimental sensitivity would be reached only for extremal black holes.
\end{abstract}

\linespread{1.2}


\vskip 1.cm

\section{Introduction}

If primordial Black Holes (BH) \,\cite{ZelNov:1967, Hawking:1971ei, Carr:1974nx} were generated in the early Universe, 
they would have emitted, via their evaporation mechanism \cite{Hawking:1974sw}, not only the Standard Model  (SM) particles, 
but also all existing particles beyond the SM with mass below their Hawking temperature.
It was soon proposed that such particles might be responsible for the excess of baryons over anti-baryons \cite{Hawking:1974rv, Zeldovich:1976vw},
that they might account for some or all of the Dark Matter (DM) we observe today \cite{Fujita:2014hha, Lennon:2017tqq, Morrison:2018xla, Masina:2020xhk}, and that they might even provide a contribution to Dark Radiation (DR) \cite{Lennon:2017tqq, Hooper:2019gtx, Lunardini:2019zob, Hooper:2020evu}.

The masses of the primordial BHs could be in the broad range $10^{-5}-10^9$ g, {\it i.e.} from the Planck mass up to the mass allowing for evaporation before the nucleosynthesis epoch.
Apart from the case of gravitino production \cite{Khlopov:2004tn, Khlopov:2008qy}, the primordial BH density at formation for the range $10^{-5}-10^9$ g is at present unconstrained, as reviewed \emph{e.g.} in ref.\,\cite{Carr:2020gox}. 
There however upper bounds\,\cite{Papanikolaou:2020qtd, Domenech:2020ssp} on the fraction of the universe collapsed into primordial BHs  from possible backreaction gravitational waves. 
Depending on the fraction of primordial BHs at formation with respect to radiation, $\beta$, there is the possibility that the Universe was radiation or BH dominated at the evanescence of the BHs \cite{Barrow:1990he, Baumann:2007yr, Fujita:2014hha}: this situation is referred to as radiation or BH domination, respectively. 

Much work has been done in the past, and also recently, in the case of non-rotating, {\it i.e.} Schwarzschild BHs, for both DM and DR. 
It is natural to ask what changes in the case of rotating, {\it i.e.} Kerr BHs \cite{Kerr:1963ud, Page:1976ki}: this is the goal of the present work.
To motivate the interest in the extension to the Kerr case, let us first summarize the results so far obtained for DM and DR in the Schwarzschild case.

As for DM, Fujita \emph{et al.} \cite{Fujita:2014hha}, assuming BHs domination,
found that a significant contribution could come from stable particles that are either superheavy or light, that is with masses in the MeV range. 
In the light case, DM candidates would be warm, while in the superheavy case they would be cold. 
Exploiting the warm DM velocity constraints available at that time \cite{Viel:2005qj}, ref.\,\cite{Fujita:2014hha} discussed the lower limits on the mass of the light DM candidates, using an order-of-magnitude argument (for an up-to-date presentation, see \cite{Masina:2020xhk}).
More sophisticated analysis were done in refs. \cite{Lennon:2017tqq, Baldes:2020nuv}.
Ref. \cite{Auffinger:2020afu} presents a complete study on the viability of warm DM candidates from the evaporation of primordial BHs:
it was found that, assuming BH domination, the scenario of warm DM is excluded for all spins and for all BH masses in the range $10^{-5}-10^9$ g;
for radiation domination, upper limits on $\beta$ (or, equivalently, on the warm DM mass) were derived for the various DM spins.

It is natural to ask what happens to warm DM in the Kerr case \cite{Kerr:1963ud, Page:1976ki}. 
Since Kerr BHs have shorter lifetimes, one might expect that the mean velocity of DM gets reduced, and the tension with structure formation alleviated. 
This is what we study in the next sections, finding that the tension with structure formation is practically unchanged for DM particles with spins $s=0,1/2,1$, while it is even exacerbated for $s=2$ increasing the value of the Kerr BH spin.

As for DR, in the Schwarzschild case, Hooper {\it et al.}\,\cite{Hooper:2019gtx} pointed out that, for BH domination, the contribution to the effective number of relativistic degrees of freedom, $\Delta N_{eff}$, might be at hand of future observations, for DR particles with spins $s=0,1/2,1$, but not for $s=2$.
The effect of primordial BHs merging was recently reconsidered in ref.\,\cite{Hooper:2020evu}, showing that a population of Kerr BHs might be formed and that their evaporation would produce a significant fraction of hot (high-energy) gravitons: exploiting previous results from \cite{Fishbach:2017dwv},
ref.\,\cite{Hooper:2020evu} finds an increase in $\Delta N_{eff}$ at a potentially observable level, even for non extremal BH spin values  (such as $a_*=0.7$), for BHs evaporating just before nucleosynthesis. 

Here we reconsider the calculation of $\Delta N_{eff}$ in the Kerr case, for any value of the BH spin and independently of the mechanism responsible for the rotation. In order to numerically account for the greybody factors associated to the different spins, we use the recently developed and publicly available code \texttt{BlackHawk} \cite{Arbey:2019mbc}. We find less optimistic results than those of ref.\,\cite{Hooper:2020evu}.
According to our results, the contribution to $\Delta N_{eff}$ by hot massless gravitons, would escape detection by projected experiments, unless in the case of extremal BH spin values, for all the range of BH masses (that is $10^{-5}-10^9$ g).

In this work we consider BH evaporation as the only production mechanism. The consequences of allowing for other production mechanisms have been recently explored in refs.\,\cite{Gondolo:2020uqv} and \cite{Bernal:2020ili, Bernal:2020bjf}. 
For an updated analysis of the possibility that the matter-antimatter asymmetry is due to particles produced by primordial BHs evaporation, 
we refer the interested reader to ref.\,\cite{Hooper:2020otu} for GUT baryogenesis, and to ref.\,\cite{Perez-Gonzalez:2020vnz} for leptogenesis. 

The paper is organized as follows. In sec.\,2 we introduce our notation and review the basics for primordial Kerr BHs. In sec.\,3, the evaporation process for Kerr BHs is discussed. In sec.\,4 we show the dynamics of the relevant densities from formation to evaporation, while in sec.\,5
we calculate the distribution of the momentum of the emitted particle at the time of evaporation, for the various particle spins. The number and energy densities at evaporation are calculated in sec.\,6, and the upper limit on the DR mass is reviewed in sec.\,7. The contributions to DM and DR are studied in secs.\,8 and 9 respectively.
We draw our summary in the conclusive sec.\,10. 

In order to have a better control of our formulas for dimensional analysis and numerical computations, we do not use natural units.

\section{Preliminaries on primordial BHs}  

The formation from early Universe inhomogeneities of collapsed objects\,\cite{ZelNov:1967}, later called primordial BHs, was considered in refs.\,\cite{Hawking:1971ei, Carr:1974nx, Hawking:1974sw}. 
Since inflation removes all pre-existing inhomogeneities, any cosmologically interesting primordial BH density has however to be created after inflation. 
We refer to\,\cite{Carr:2020gox, Khlopov:2013ava} for reviews of the various mechanism proposed, with proper references to the associated literature.
In the following, we present our notations and review the relevant formulas for primordial Kerr BHs formation and dynamics.

\subsection{Radiation dominated era}

According to the first Friedmann equation (neglecting the curvature and cosmological constant terms), 
the early Universe evolution is described by
\beq
 \left( \frac{\dot a}{a} \right)^2   \equiv H(t)^2 =\frac{8 \pi G}{3 }   \rho(t)\,,
\label{eq-F1}
\eeq
where $a(t)$ is the scale factor, $H(t)$ is the Hubble parameter, $\rho(t)$ is the mass density of the Universe 
and $G$ is the Newton gravitational constant, $G \simeq 6.674 \times 10^{-11} \,\rm{ m^3 / (kg \,s^2)}$.
The Planck mass is $M_{Pl}= \sqrt{ \hbar c / G } \approx 1.221 \times 10^{19}\, \rm{GeV}/c^2  \approx  2.176 \times 10^{-8} $\,kg. 

In the early hot and dense Universe, it is appropriate to assume an equation of state corresponding to a fluid of radiation (or relativistic particles). 
During radiation domination, $\rho \propto a^{-4}$, $a(t) \propto t^{1/2}$, and
\beq
H(t)= \frac{1}{2 t }\,\,.
\label{eq-raddom}
\eeq
At relatively late times, non-relativistic matter eventually dominates the mass density over radiation. 
A pressureless fluid 
leads to the expected dependence $\rho \propto a^{-3}$, $a(t) \propto t^{2/3}$, and
\beq
H(t) = \frac{2}{3 t} \,.\,
\label{eq-matdom}
\eeq

The radiation mass density (at high temperatures) can be approximated by including only those particles which are in thermal equilibrium and have masses below the temperature $T_R$ of the radiation bath:
\beq
 \rho_R(t) = \frac{\pi^2 g_*(t)  } {30}  \frac{(k_B T _R(t))^4 }{(\hbar \,c)^3 \,c^2} \,\,\, , \,\,\, \,\,
 g_*(t)= \sum_B g_B + \frac{7}{8}\sum_F g_F \,\,\, , 
 \label{eq-rhoRT}
\eeq
where $k_B$ is the Boltzmann constant, $k_B \simeq 8.617 \times 10^{-5} $ eV/K, 
$\hbar$ is the reduced Planck constant, $\hbar=6.582 \times 10^{-16}$ eV s, 
$c$ is the velocity of light in vacuum, $c=2.998 \times 10^{8}$ m/s,
and $g_{B(F)}$ is the number of degrees of freedom (dof) of each boson (fermion). 

Below the electron mass, 
only the photon ($g_\gamma=2$) and three light left-handed neutrinos contribute, 
so that $g_*(t)=7.25$. 
Below the muon mass, 
also the electron (and the positron) has to be included, 
so that $g_*(t)=10.75$. 
For the full SM, here defined including three light left-handed neutrinos, $g_*(t)= 106.75$. 
At higher temperatures, $g_*(t)$ will be model-dependent. 
Including the (massless) graviton, ($g_G=2$), has the effect of adding $2$ units to the previously mentioned values of $g_*(t)$.

\subsection{Formation of primordial BHs}

As reviewed for instance in ref.\,\cite{Carr:2020gox}, if a primordial BH forms at the time $t_f$ during the radiation dominated era,
typically its mass is close to the value enclosed by the particle horizon near the end of inflation:
\beq
M_{BH} =  \gamma   \frac{4 \pi}{3}  \rho_R(t_f) \left( 2\, c \,t_f \right)^3 =\gamma   \frac{4 \pi}{3}  \rho_R(t_f)  \left( \frac{ c }{H(t_f)}   \right)^3\,,
\label{eq-MPBH}
\eeq
where $\gamma \lesssim 1$ is a numerical factor that depends on the details of the gravitational collapse, 
$\rho_R(t_f)$ and $H(t_f)$ are respectively the radiation density and the Hubble parameter at the formation of the BH, 
and in the last equality we used eq.\,(\ref{eq-raddom}). 
Using eq.\,(\ref{eq-F1}), we can also write 
\beq
 M_{BH} =   \frac{\gamma}{2} \frac{(M_{Pl}c^2) ^2}{ \hbar \,H(t_f)}\frac{1}{c^2} \approx  \gamma   \frac{10^{10} \,\rm{GeV}} {\hbar \,H(t_f)}  10^4 \,\rm{g}  \gtrsim \frac{\gamma}{3} \,{\rm  g} \, ,
\label{eq-PBH}
\eeq
where the last lower bound follows from the fact that CMB observations put a naive upper bound (which does not apply to all inflationary models) on the Hubble scale during inflation, $\hbar H_I \lesssim 3 \times 10^{14}$ GeV at $95\%$ C.L. \cite{Akrami:2018odb}, and $H(t_f) \lesssim H_I$.
In the literature the value $\gamma=1/(3\sqrt{3})\approx 0.2$ is usually taken as reference value\,\cite{Carr:2020gox}; in this case the lower limit would become $M_{BH} \gtrsim 0.07$\,g. 
In any case, the BH mass should be larger than the Planck mass, $M_{BH}\gtrsim 10^{-5}$\,g.
As is well known, there is also an upper bound, $M_{BH} \lesssim 10^9$\,g, 
because the primordial BH evaporation products have negative effects on nucleosynthesis, see e.g. ref.\,\cite{Keith:2020jww}. 
The range of primordial BH masses between these bounds is at present generically unconstrained\,\cite{Carr:2020gox}. 

Recalling eq.\,(\ref{eq-raddom}), the primordial BHs formation time is easily found from eq.\,(\ref{eq-PBH}) to be
\beq
\frac{t_{f}}{\hbar}=\frac{1} {\gamma}  \frac{M_{BH} c^2}{ (M_{Pl} c^2)^2  } \,.
\label{eq-tfBH}
\eeq
As for the radiation temperature at formation,
combining eqs.\,(\ref{eq-rhoRT}), (\ref{eq-F1}) and (\ref{eq-PBH}), we have 
\beq
k_B T_R(t_{f})  
 = \left(  \frac{45 \gamma^2  } {16 \pi^3 g_*(t_{f}) }   \right)^{1/4}  \left( \frac{M_{Pl}  }{M_{BH}} \right)^{1/2}\, M_{Pl} c^2 \,.
 \label{eq-Tf}
\eeq
The temperature and the time at formation of primordial BHs, as a function of the their mass at formation, are plotted {\it e.g.} in fig.\,1 of ref.\,\cite{Masina:2020xhk}.

It is useful to introduce the parameter $\beta$ defined as the BH over the radiation mass density at the formation time
\beq
\beta= \frac{\rho_{BH}(t_f)}{\rho_R(t_f)}= M_{BH} \frac{n_{BH}(t_f)}{\rho_R(t_f)} \, ,
\label{eq-defbeta}
\eeq
where $n_{BH}(t_f)$ is the primordial BH number density at formation, whose
explicit expression can be obtained by combining eqs.\,(\ref{eq-defbeta}) and (\ref{eq-MPBH}) 
with eq.\,(\ref{eq-F1}), so that 
\beq
n_{BH} (t_f) 
= \beta \, \gamma^2 \, \frac{3}{32 \pi} \frac{(M_{Pl} c^2)^6}{(M_{BH} c^2)^3 (\hbar c)^3} \,\,.
\label{eq-nBHtf}
\eeq

\subsection{Kerr primordial BHs } 

Let $J$ be the angular momentum 
of the rotating (uncharged) Kerr BH.
While the mass of the BH can take any positive value, 
there is an upper limit on $J$, 
\beq
\frac{J}{\hbar} <  \frac{ M_{BH}^2}{M_{Pl}^2}
\,\, .
\eeq
This suggest to define a dimensionless spin parameter $a_*$, 
\beq
a_*= \frac{J}{\hbar }\frac{ M_{Pl}^2}{M_{BH}^2} 
\,\, ,
\eeq 
such that $0<a_*<1$. Extremal Kerr BHs are characterized by having $a_*$ very close to $1$, let say $a_*>0.9$ for definiteness.

The size of a Schwarzschild ($a_*=0$) BH, as determined by the radius of the event horizon, is proportional to its mass through
\beq
r_S =2 \hbar c \, \frac{ M_{BH} c^2}{ (M_{Pl} c^2)^2 } \,\, .
\eeq
For a BH with nonzero angular momentum, the radius of the event horizon, $r_+$, is smaller than the Schwarzschild radius
\beq 
\frac{r_+}{r_S} =  \frac{ 1+ \sqrt{ 1-a_*^2} }{2}    \,\,\, ,
\eeq
until an extremal BH could have an event horizon close to $r_S/2$.

\section{Evaporation of Kerr primordial BHs }

Here we review the basic formulas describing the evaporation of a Kerr BH \cite{Kerr:1963ud, Page:1976ki},  
following the notations of refs.\,\cite{Arbey:2019jmj, Arbey:2019mbc}.

Consider a Kerr BH of mass $M_{BH}(t)$ (we neglect the time dependence only when we refer to the formation time). 
The Hawking radiation mimics thermal emission from a blackbody with a temperature $T_{BH}(t)$, 
given by\,\cite{Hawking:1974sw, Page:1976ki,Page:1976df} 
\beq
k_B \,T_{BH}(t)
= \frac{ 1}{8 \pi }   \frac{(M_{Pl} c^2)^2 }{M_{BH}(t) c^2}  \, \frac{2}{ 1 + \frac{1}{ \sqrt{1- a_*(t)^2}} } 
 \,\,.
\label{eq-TBH}
\eeq
In the following, we denote by $T_{BH}$ the Hawking temperature at formation, $T_{BH}=T_{BH}(t_f)$.
For the Schwarzschild case, for which $a_*=0$, the formation temperature is denoted by $T^S_{BH}$.

As discussed e.g. in \cite{Page:1976ki}, at the time $t$, such a hole emits particles of type $i$, spin $s_i$ and total energy between $(E, E+dE)$ at a rate, per degree of freedom (dof), given by
\beq
\frac{1}{g_i} \frac{d^2N_i}{dt \,dE} =   \frac{d^2N}{dt \,dE}
=  \frac{1}{2 \pi \hbar } \sum_{\ell,m} \Gamma_{s_i \ell m}(E,M_{BH}(t),a_*(t)) 
\frac{1}{e^\frac{E'}{k_B T_{BH}(t)} -(- 1)^{2 s_i}} \,\, ,
 \label{eq-d2NdtdE}
 \eeq
where $E'=E-m \Omega$ is the total energy of the particles, 
taking into account the BH horizon rotation velocity on top of the total energy $E^2= p^2 c^2+ m_i^2 c^4$ (where $m_i$ is the particle mass),
$m$ is the particle angular momentum projection $m \in [-\ell, +\ell]$, and
$\Omega= \hbar c \,a_*   /(2 r_+  ) $. 
The greybody factors $\Gamma_{s_i \ell m}$ are dimensionless absorption probabilities for the emitted species 
(in general functions of $E$, $M_{BH}(t)$, $a_*(t)$ and the particle's internal dof and rest mass),
and $g_i$ are the internal dof of the $i$-th particle, which account for the polarization and color dof.

Let us consider in some detail the SM. For the Higgs boson ($s=0$), $g_{h^0}=1$. For the massless ($s=1$) photon and the 8 gluons, $g_\gamma =2$ and  $g_g =16$.
For the massive ($s=1$) W$^\pm$ and Z, $g_{W^+} = g_{W^-} =g_{Z} = 3$. As for the fermions ($s=1/2$): the charged leptons, being Dirac fermions, have
$g_e=g_\mu=g_\tau =4$; the neutrinos have $g_{\nu_e}=g_{\nu_\mu}=g_{\nu_\tau}=2(4)$ in the case they are Majorana (Dirac) particles, respectively;
the quarks have $g_u=g_c=g_t=g_d=g_s=g_b=12$. Finally, one might also include the massless graviton ($s=2$), with $g_G=2$.

\subsection{Rate of mass and angular momentum loss }

The rate of mass loss for an evaporating BH is proportional to the total power emitted
\beq
\frac{c^4 M_{Pl}^4}{\hbar} \frac{f(M_{BH},a_*)}{M_{BH}^2} 
=-c^2 \frac{dM_{BH}}{dt} =  \frac{dE}{dt} 
= \sum_i  \int_0^\infty  dE \, E\,\frac{d^2N_i}{dt\, dE} \,\, 
\,\,.
\label{eq-Mloss}
\eeq
where, to parametrize this, Page \cite{Page:1976df,Page:1976ki} introduced the adimensional Page function, $f(M_{BH},a_*)$.

For the evolution of the BH angular momentum, Page introduced the adimensional function $g(M_{BH},a_*)$ such that
\beq
M_{Pl}^2 c^2\, g(M_{BH},a_*) = -\frac{M_{BH}}{a_*} \frac{dJ}{dt}  = \frac{M_{BH}}{a_*} \sum_i \int dE\, j_i\, \frac{d^2N_i}{dt dE}  \,\, ,
\eeq
where $j_i$ is the angular momentum taken by the $i$-th particle, $j_i = m \hbar$. 
The time dependence of $M_{BH}$ and $a_*$ are left understood in the above equations. 

Inverting these equations and using the definition of $a_*$, one obtains the differential equations governing the mass
and spin of a Kerr BH
\beq
\frac{dM_{BH}}{dt} =-    \frac{c^2 M_{Pl}^4}{\hbar}  \frac{f(M_{BH},a_*)}{M_{BH}^2} \,\, ,
\label{eq-fK}
\eeq
\beq
\frac{da_*}{dt} 
= a_*  \frac{ (M_{Pl}  c^2)^4}{\hbar (M_{BH} c^2)^3}  \left(2 f(M_{BH},a_*) - g(M_{BH},a_*) \right) \,\, .
\label{eq-gK}
\eeq

\begin{figure}[h!]
\vskip .0 cm 
 \begin{center}
 \includegraphics[width=7.7cm]{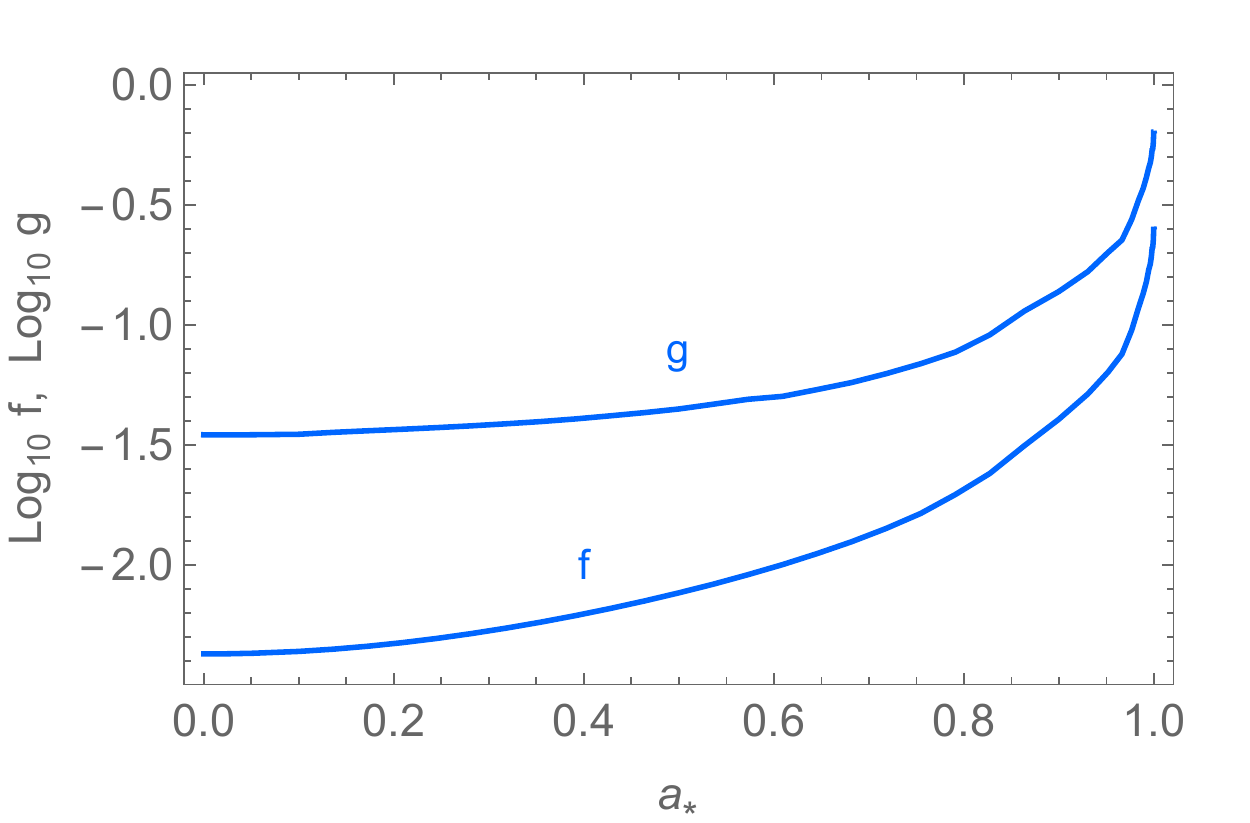}  \,\, \includegraphics[width=7.6cm]{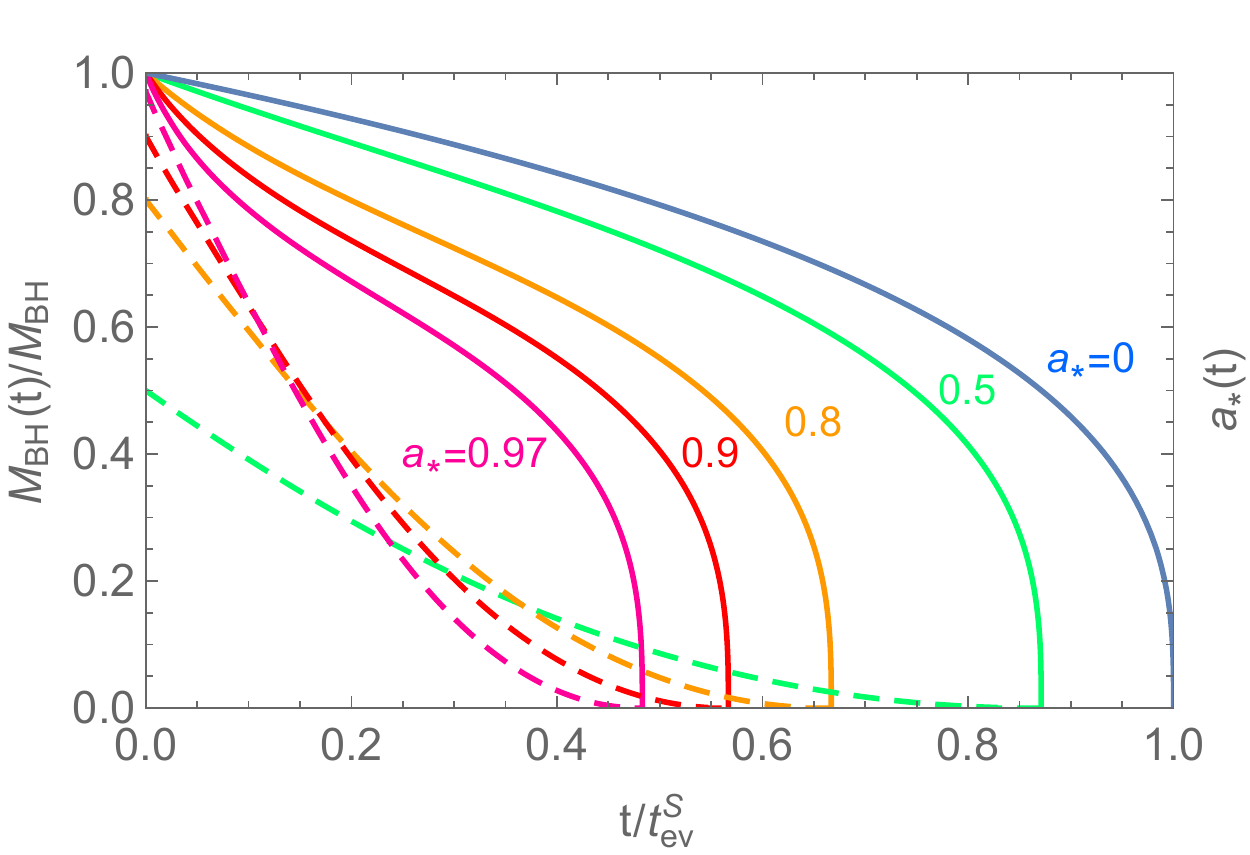}  
 \end{center}
\caption{\baselineskip=15 pt \small  
Left: The Page functions $f$ and $g$ as a function of $a_*$, within the SM (including gravitons) and for BHs with initial mass $M_{BH}\lesssim 10^{10}$ g.
Right: Time dependence of the mass (solid) and spin (dashed) of a Kerr BH in the SM (including gravitons), for $a_*=0, 0.5, 0.8, 0.9,0.97$ and for any value of $M_{BH}\lesssim 10^{10}$ g.  }
\label{fig-K}
\vskip .2 cm
\end{figure}

For the SM (including gravitons), the functions $f(M_{BH},a_*)$ and $g(M_{BH},a_*)$ are constant over the whole range of BH masses we are interested in, namely $(10^{-5}-10^9)$ g, while their dependence on the spin parameter $a_*$ is shown in the left panel of fig.\,\ref{fig-K}. 
We calculated the $f(M_{BH},a_*)$ and $g(M_{BH},a_*)$ Page functions using the code BlackHawk\,\cite{Arbey:2019mbc}, and
we checked that they are consistent with previous numerical estimates\,\cite{Page:1976ki,Dong:2015yjs,Taylor:1998dk}. 
In particular, in the Schwarzschild case, one has $f_S(M_{BH})=f(M_{BH},0) = 4.27 \times 10^{-3}$.

\subsection{BH lifetime}

Given any initial values for the BH mass and spin, the functions $f(M_{BH},a_*)$ and $g(M_{BH},a_*)$ allow one to obtain the time evolution of the BH mass and spin, by using eqs. (\ref{eq-fK}) and (\ref{eq-gK}). 

For a Schwarzschild BH, the lifetime equals the time of the BH evanescence, $t_{ev}^S$, so that
\beq
\frac{t_{ev}^S}{\hbar} = \frac{1}{c^2 M_{Pl}^4 }  \int_0^{M_{BH}} dM  \frac{M^2}{f(M,0)}
= \frac{1}{3  f_S(M_{BH})} \frac{(M_{BH} c^2)^3}{(M_{Pl} c^2)^4} \,\, ,
\label{eq-tevBH}
\eeq
where in the last equality we assumed $f(M,0)$ to be constant over the BH lifetime (as in the SM). 
For Schwarzschild, one thus obtains a simple time dependence for the ratio of the BH mass at time $t$ over its initial value, 
\beq
\frac{M^S_{BH}(t)}{M_{BH} } = \left( 1- \frac{t}{t_{ev}^S}\right)^{1/3} \, ,
\label{eq-MBHt}
\eeq
as illustrated in the right panel of fig.\,\ref{fig-K}. 

For the more general Kerr case, considering for definiteness the SM (including gravitons), we display in fig.\,\ref{fig-K} the time dependence of the mass ratio $M_{BH}(t)/M_{BH}$ (solid lines) and of the spin $a_*(t)$ (dashed lines). In particular, we select some representative initial spin values, $a_*=0.5,0.8,0.9,0.97$. 
The results hold for any value of $M_{BH}\lesssim 10^{10}$ g. 
The lifetime is more and more reduced the higher is the initial spin of the BH:
this means that the evaporation time in the Kerr case is smaller than in the Schwarzschild case. To account for the shortening in the lifetime, it is useful to introduce the parameter $\alpha_K$ such that
\beq
t_{ev} = \alpha_K  t_{ev}^S \,\, .
\label{eq-aK}
\eeq

From fig.\,\ref{fig-K}, one might suppose that a Kerr BH behaves like a Schwarzschild BH in the final stages of its lifetime.
In particular, in his last period, the time dependence of a Kerr BH mass is similar to the one of a Schwarzschild BH 
with an initial effective mass given by $M^{eff}_{BH}=\alpha_K^{1/3} M_{BH}$.

It is not difficult to study the effect of adding beyond SM particles, like for instance axions and right-handed neutrinos. 
As far as the number of dof of the additional particles is small with respect to the SM one, the inclusion of such additional particles does not change appreciably the results of fig.\,\ref{fig-K}.

\section{From formation to evaporation}

\subsection{Radiation vs BH domination}

Let us define the ratio $f(t)=\rho_{BH}(t)/\rho_R(t)$. 
Since $\rho_{BH}(t) \propto 1/a(t)^3$, while $\rho_R(t)\propto 1/a(t)^4$, 
such ratio increases as the scale factor, $f(t)\propto a(t)$. 
It is thus possible that BHs come to dominate the energy content of the universe 
before they completely evaporate \cite{Barrow:1990he, Baumann:2007yr, Fujita:2014hha}:
this situation is referred to as BH domination. 
The scenario in which evaporation takes place before BH domination might occur is referred to as radiation domination. 

We define $\bar \beta$ to be the maximum value of $\beta$ 
corresponding to radiation domination, namely the value of $\beta$ leading to ${f(t_{ev})} = 1$;
this value can be obtained from the following relation,
\beq
 {\bar \beta} = \frac{f(t_f)}{f(t_{ev})} = \frac{a(t_f)}{a(t_{ev})} 
 = \left( \frac{t_f}{\alpha_K t^S_{ev}} \right)^{1/2} 
=  \frac{1}{\alpha_K^{1/2}}   \left(     \frac{ 3 f_S(M_{BH})}{\gamma}  \right)^{1/2}  \frac{M_{Pl} } {M_{BH} }  
\,\, ,
\label{eq-barbeta}
\eeq
where we used eqs.\,(\ref{eq-aK}), (\ref{eq-tfBH}) and (\ref{eq-tevBH}).
For all the values of $\beta \lesssim \bar \beta$,
the primordial BHs evaporate before they come to dominate the matter content of the Universe, 
and the increase in the scale factor is simply given by
\beq
\frac{a(t_f)}{a(t_{ev})} = {\bar \beta}  \,\, .
 \label{eq-summR}
\eeq

For all the values of $\beta \gtrsim \bar \beta$, BH domination occurs. 
We have thus to consider: first, the radiation dominated period from the formation time, $t_f$, to the time when BHs start to dominate, $t_{BH}$, such that $f(t_{BH})= 1$; and second, the matter dominated period from $t_{BH}$ to $t_{ev}$. 
The first period is characterized by the following increase in the scale factor
\beq
{\beta} = \frac{f(t_f)}{f(t_{BH})} = \frac{a(t_f)}{a(t_{BH})} = \left( \frac{t_f}{t_{BH}} \right)^{1/2} \,\, ,
\eeq
while the second period is characterized by
\beq
\frac{a(t_{BH})}{a(t_{ev})} = \left( \frac{t_{BH}}{t_{ev}} \right)^{2/3} =  \left( \frac{1}{\beta^2} \frac{t_{f}}{t_{ev}} \right)^{2/3}  
= \left( \frac{1}{\beta^2} \frac{t_{f}}{ \alpha_K t^S_{ev}} \right)^{2/3}  
 =\left( \frac{\bar \beta}{\beta}  \right)^{4/3}\,\, ,
\eeq
so that, putting together,  the total increase in the scale factor is given by
\beq
\frac{a(t_f)}{a(t_{ev})} = \frac{a(t_f)}{a(t_{BH})} \frac{a(t_{BH})}{a(t_{ev})} 
= \frac{\bar \beta^{4/3}}{\beta^{1/3}}= \bar \beta  \left( \frac{\bar \beta}{\beta}  \right)^{1/3} \,\, .
 \label{eq-summBH}
\eeq

\subsection{BH density at evaporation}

The BH number density at the time of evaporation is related to the one at formation by
\beq
n_{BH} (t_{ev}) = n_{BH} (t_f) \left( \frac{a(t_f)}{a(t_{ev})}\right)^3 \,.
\label{eq-nBHtev}
\eeq
Using the equation above together with eqs.\,(\ref{eq-nBHtf}), (\ref{eq-barbeta}) and (\ref{eq-summR}), one obtains for radiation domination
\beq
n_{BH}(t_{ev}) = \frac{1}{\alpha_K^{3/2}} \,   \beta \,\gamma^{1/2} \,\,   \, \frac{3}{ 32 \pi} \, 
  { (3 f_S(M_{BH}))^{3/2}} \,  \left( \frac{M_{Pl} } {M_{BH} }\right)^6 \,\frac{(M_{Pl} c^2)^3}{(\hbar c)^3} \, \,\, ,
 \label{eq-nBHtevR}
\eeq
while, using eq.\,(\ref{eq-summBH}), one obtains for BH domination 
\beq
n_{BH}(t_{ev}) =   \frac{1}{\alpha_K^{2}}  \, \, \frac{3}{32 \pi} \, 
  ( 3 f_S(M_{BH}) )^2  \,  \left( \frac{M_{Pl} } {M_{BH} }\right)^7 \,\frac{(M_{Pl} c^2)^3}{(\hbar c)^3} \, \,\, ,
     \label{eq-nBHtevBH}
\eeq
which displays an increase with respect to radiation domination by the factor $\bar \beta/\beta$.

\subsection{Radiation temperature at evaporation}

One can grossly assume that all the energy density stored in the BHs goes, after their evaporation, into the radiation energy density of the SM particles and of possibly beyond SM particles emitted by the BH:
\beq
\rho_{BH}(t_{f}) \left( \frac{a(t_f)}{a(t_{ev})} \right)^3 =\rho_{BH}(t_{ev}) \approx \rho_{SM}(t_{ev}) + \rho_{BSM}(t_{ev}) \,\, .
\eeq
Both the SM and beyond SM particles are emitted with a spectrum of momenta that is not thermal.
However, the SM particles produced in the evaporation of the BH rapidly thermalize as soon as they are emitted. 
In this work we assume that the beyond SM particles are stable and interact feebly or only gravitationally, so that they never come in thermal equilibrium.

In the case of radiation domination,
the radiation energy density from the SM particles emitted by the BH is negligible with respect to the radiation present since the formation of the BH,
 $ \rho_{BH}(t_{ev})  << \rho_{R}(t_{ev})$.
Combining eqs. (\ref{eq-F1}), (\ref{eq-raddom}) and eq.\,(\ref{eq-aK}), 
\beq
\frac{8 \pi G}{3} \rho_R(t_{ev})  = \frac{1}{4 t_{ev}^2}= \frac{1}{4 \alpha_K^2 {t^S_{ev}}^2} \,\, ,
\eeq
and using eqs.\,(\ref{eq-rhoRT}) and (\ref{eq-tevBH}), one obtains
\beq
k_B T_R(t_{ev}) = \frac{1}{\alpha_K^{1/2}} \,
 \left(   3 f_S(M_{BH}) \right)^{1/2}  
 \left( \frac{ 45  }{16 \pi^3 g_{*}(t_{ev})}     \right)^{1/4}  \left( \frac{     M_{Pl} }{  M_{BH}  }  \right)^{3/2} (M_{Pl} c^2) 
 \approx  \frac{1}{\alpha_K^{1/2}} \, k_B T^S_R(t_{ev})\,\, ,
\label{eq-TevR} 
\eeq
namely that the temperature in the Kerr case is slightly higher than in the Schwarzschild case:
notice indeed that, since $\alpha_K$ is in general bigger than $1/2$ (see right panel of fig.\,\ref{fig-K}), the function $g_{*}(t_{ev})$ is very similar to what it is in the Schwarzschild case, here denoted by $g_{*}(t^S_{ev})$. The values of $g_{*}(t^S_{ev})$, as a function of the BH mass, are shown in the right panel of fig. 3 of ref.\,\cite{Auffinger:2020afu}.

In the case of BH domination, by definition, $\rho_{BH}(t_{ev})  >> \rho_{R}(t_{ev})$. 
If the fraction of beyond SM particles is not enormous, one also has $\rho_{SM}(t_{ev}) \approx \rho_{BH}(t_{ev})$. 
The beyond SM particles remain non thermal if they interact feebly or only gravitationally.
At the contrary, the SM particles produced in the evaporation of the BH rapidly thermalize as soon as they are emitted, so that
the radiation energy density (and thus the radiation temperature) gets a sudden increase, going up to $\rho_R(t_{ev}) \approx \rho_{BH}(t_{ev})$ after thermalization.
Assuming the period of matter domination by BHs is respectively short or long with respect to the first period of radiation domination, and combining eqs. (\ref{eq-F1}) and (\ref{eq-matdom}), we have
\beq
\frac{8 \pi G}{3} \rho_{R}(t_{ev})
=\left( 1\, {\rm or}\, \frac{16}{9} \right) \frac{1}{4 t_{ev}^2} \,\, .
\eeq
For a short or long period of matter domination by BHs, the radiation temperature after evaporation gets slightly enhanced with respect to radiation domination, by a factor going from $1$ up to $({16}/{9})^{1/4}\approx 1.15$.

\subsection{BH number to entropy density at evaporation}

For later convenience it is useful to introduce $Y_{BH}(t)$, the number to entropy density of BHs at time $t$ 
\beq
Y_{BH}(t) =  \frac{n_{BH}(t)}{s(t)} \, ,
\label{eq-YBH}
\eeq
where the entropy density is defined as 
\beq
s(t) = \frac{ 2 \pi^2 g_{*,S}(t) }{45} \frac{(k_B T_R(t))^3}{(\hbar c )^3} \,.
\label{eq-s}
\eeq
The difference between $g_{*}(t)$ and $g_{*,S}(t)$ can in general be neglected. 

For radiation domination, using eqs.\,(\ref{eq-nBHtevR}) and (\ref{eq-TevR}), 
we have that the number-to-entropy density in the Kerr and Schwarzschild cases are equal
\beq
Y_{BH}(t_{ev}) =\beta \,\gamma^{1/2} \,\,   \, \frac{3}{ 4 } \,  
 \left( \frac{ 45  }{16 \pi^3 g_{*}(t_{ev})}     \right)^{1/4}   \left( \frac{M_{Pl} } {M_{BH} }\right)^{3/2} 
 \,\, ,
 \label{eq-YBHR}
\eeq
as the dependence on $\alpha_K$ disappears.
For a short period of matter domination by BHs, using eqs.\,(\ref{eq-nBHtevBH}) and (\ref{eq-TevR}) 
\beq
Y_{BH}(t_{ev}) = \,  \, 
\frac{1}{\alpha_K^{1/2}}    ( 3 f_S(M_{BH}) )^{1/2}  \,   \, 
\, \frac{3}{4 }\, \left( \frac{ 45  }{16 \pi^3 g_{*}(t_{ev})}     \right)^{1/4} \,\left( \frac{M_{Pl} } {M_{BH} }\right)^{5/2} \,\, ,
 \label{eq-YBHBH}
\eeq
which displays an increase with respect to radiation domination by the factor $\bar \beta/\beta$.

\section{Momentum distribution at evaporation}

Eq. (\ref{eq-d2NdtdE}) gives the instantaneous spectrum of the particles of type $i$ emitted by a single BH.
The maximum of the energy distribution is at about $E \sim k_B T_{BH}$.
If the particle is sufficiently light, as we are going to assume in the following, the ultra relativistic limit applies, $E \approx c p > m_i c^2$.

The distribution of the momentum at evaporation for the particle of type $i$, normalized per dof, is a superposition of all the instantaneous distributions, each redshifted appropriately from its time of emission $t_{em}$ (see \emph{e.g.} ref.\,\cite{Bugaev:2000bz})
\beq
\frac{1}{g_i} \frac{dN_i}{d(cp)}(t_{ev})
=\int_{t_{em}}^{t_{ev}} dt \,  \frac{d^2 N}{ dt\,d(c p(t)) }  \left( \underbrace{c p(t_{ev}) \frac{a(t_{ev})}{a(t)} }_{cp(t)}, T_{BH}(t) ,a_*(t)\right) \,\,  \frac{a(t_{ev})} {a(t)} \, .
\label{eq-Fev}
\eeq
Notice that $t_{em}$ might be larger than $t_f$ if the initial BH temperature is smaller than the particle mass but, 
since we are interested in light DM, $t_{em}= t_f$.

For radiation domination, the ratio of scale factors to be put in eq.\,(\ref{eq-Fev}) is
\beq
 \frac{a(t_{ev})}{a(t)} = \left( \frac{t_{ev}}{t} \right)^{1/2} \,.
 \eeq

For BH domination, the integral of eq.\,(\ref{eq-Fev}) should be split into two contributions, corresponding to a first period of radiation domination, and a second of BH domination.
For the second period of BH domination, starting at $t_{BH}=t_f/\beta^2$ and ending at $t_{ev}$, 
the ratio of scale factors to be put in the integrand is
\beq
 \frac{a(t_{ev})}{a(t)} = \left( \frac{t_{ev}}{t} \right)^{2/3} \, \,,
\eeq
while for the first period of radiation domination, starting at $t_f$ and ending at $t_{BH}$,
the ratio of scale factors to be put in the integrand is rather
\beq
 \frac{a(t_{ev})} {a(t)} =  \frac{a(t_{ev})} {a(t_{BH})} \frac{a(t_{BH})} {a(t)} 
 = \left( \frac{t_{ev}}{t_{BH}} \right)^{2/3} \left( \frac{t_{BH}}{t} \right)^{1/2}  \, .
\eeq
For short (long) BH domination, clearly the dominant contribution comes from the first (second) period.

It is useful to define the adimensional momentum 
\beq
x(t_{ev}) \equiv \frac{c p(t_{ev})}{k_B T^S_{BH} } \, ,
\eeq
with the related adimensional momentum distribution at evaporation
\beq
\tilde F_{s_i}(x(t_{ev}))  \equiv \frac{(k_B T^S_{BH})^3}{(M_{Pl} c^2)^2} \frac{1}{g_i} \frac{dN_i}{d(cp)}(t_{ev})
\,\, ,
\label{eq-tFev}
\eeq
which has the nice property of depending only on the particle spin $s_i$ and the BH spin $a_*$, while being independent of the BH mass.

\begin{figure}[h!]
\vskip .0 cm 
 \begin{center}
\includegraphics[width= 7.6 cm]{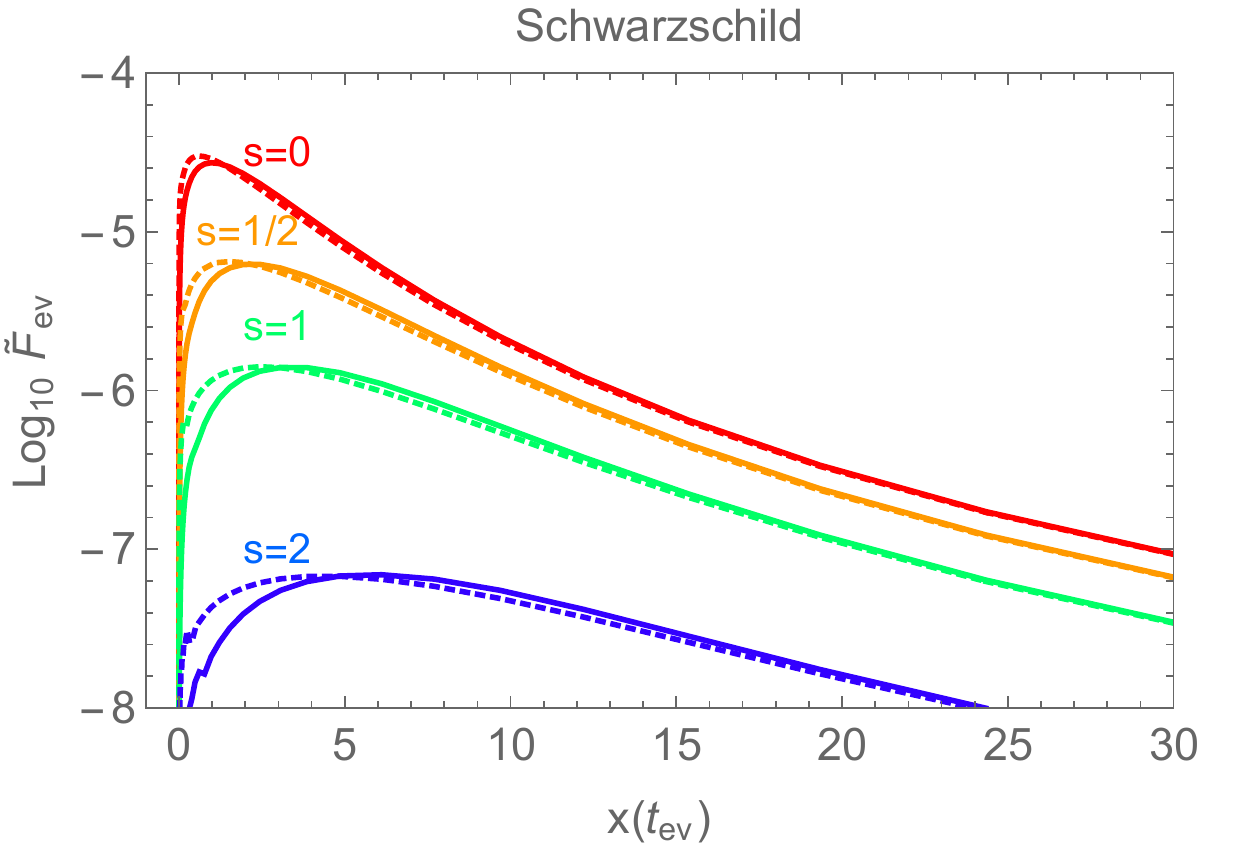}   \,\,
  \includegraphics[width= 7.6 cm]{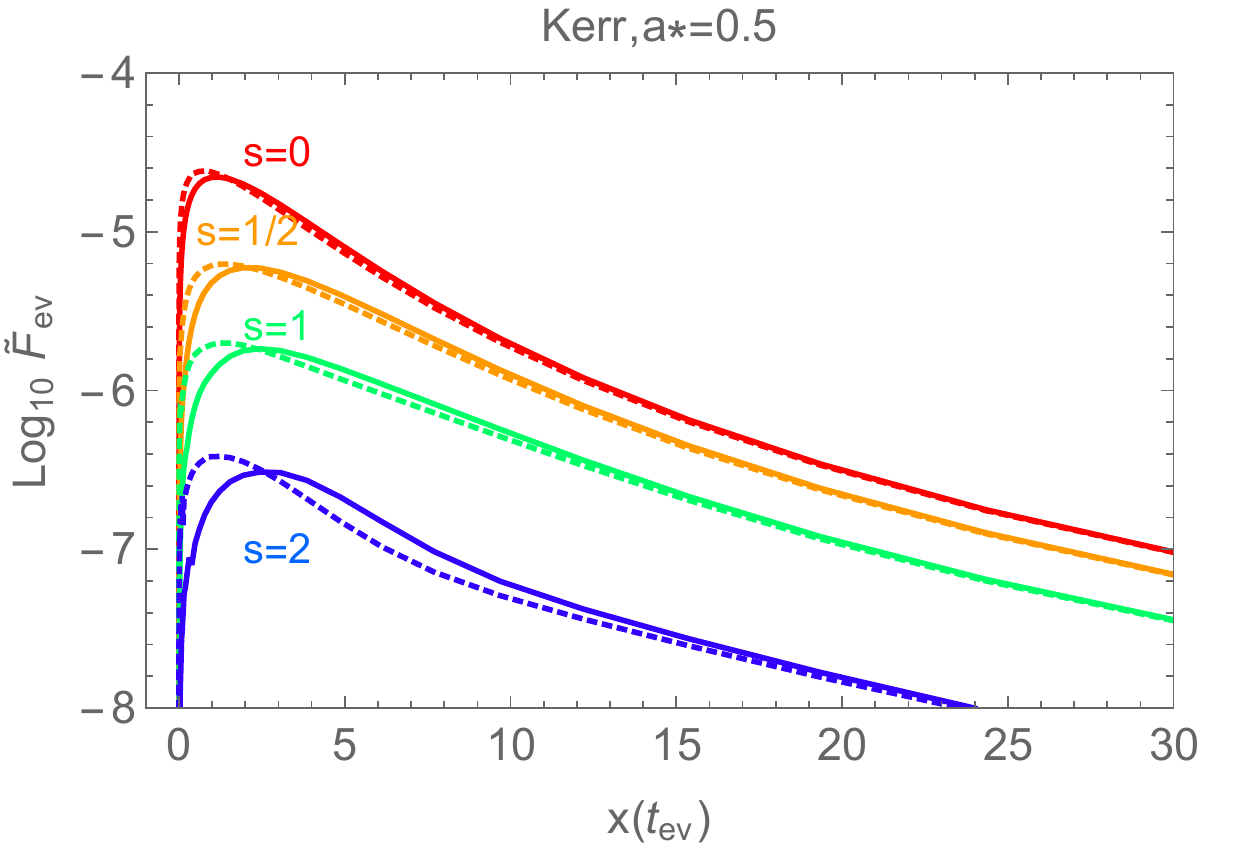}   \,\, 
 \vskip .5 cm  \includegraphics[width= 7.6 cm]{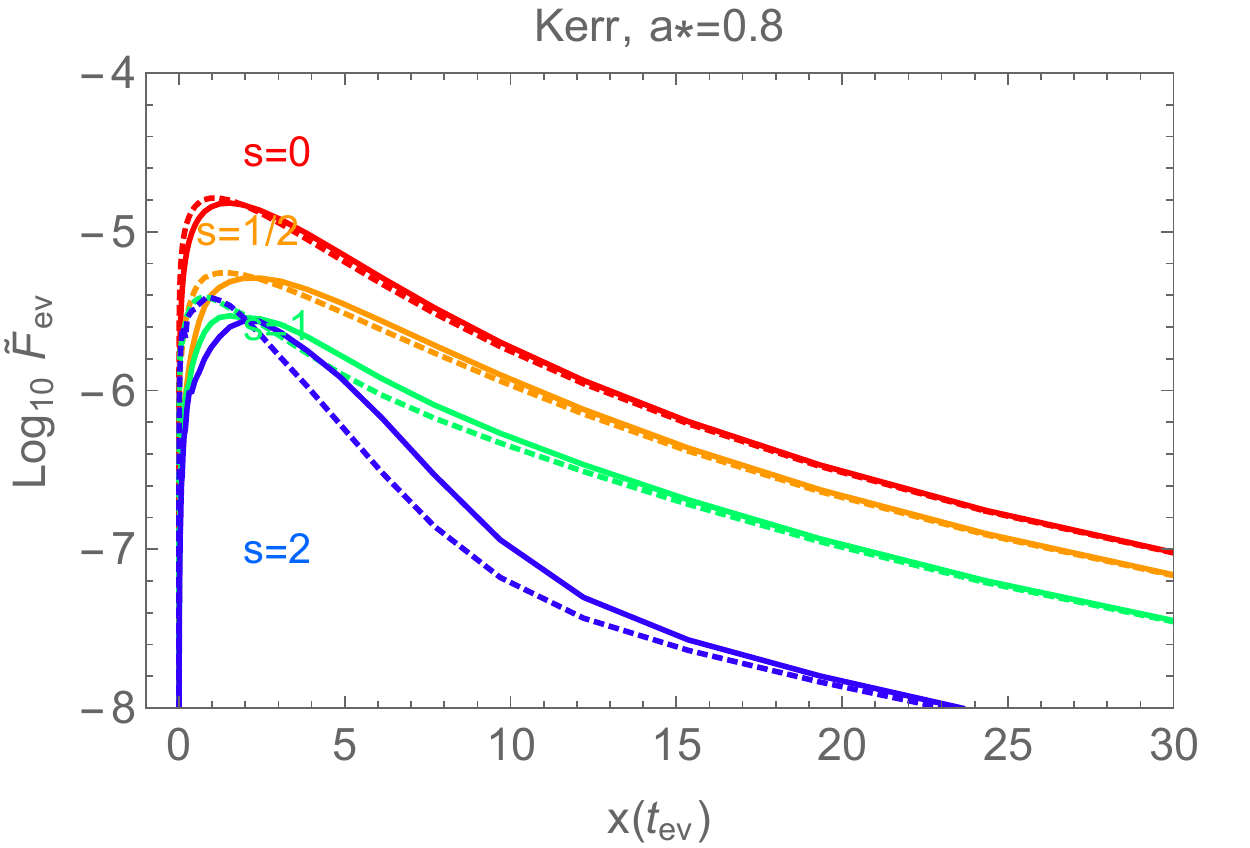}  
 \,\,  \includegraphics[width= 7.6 cm]{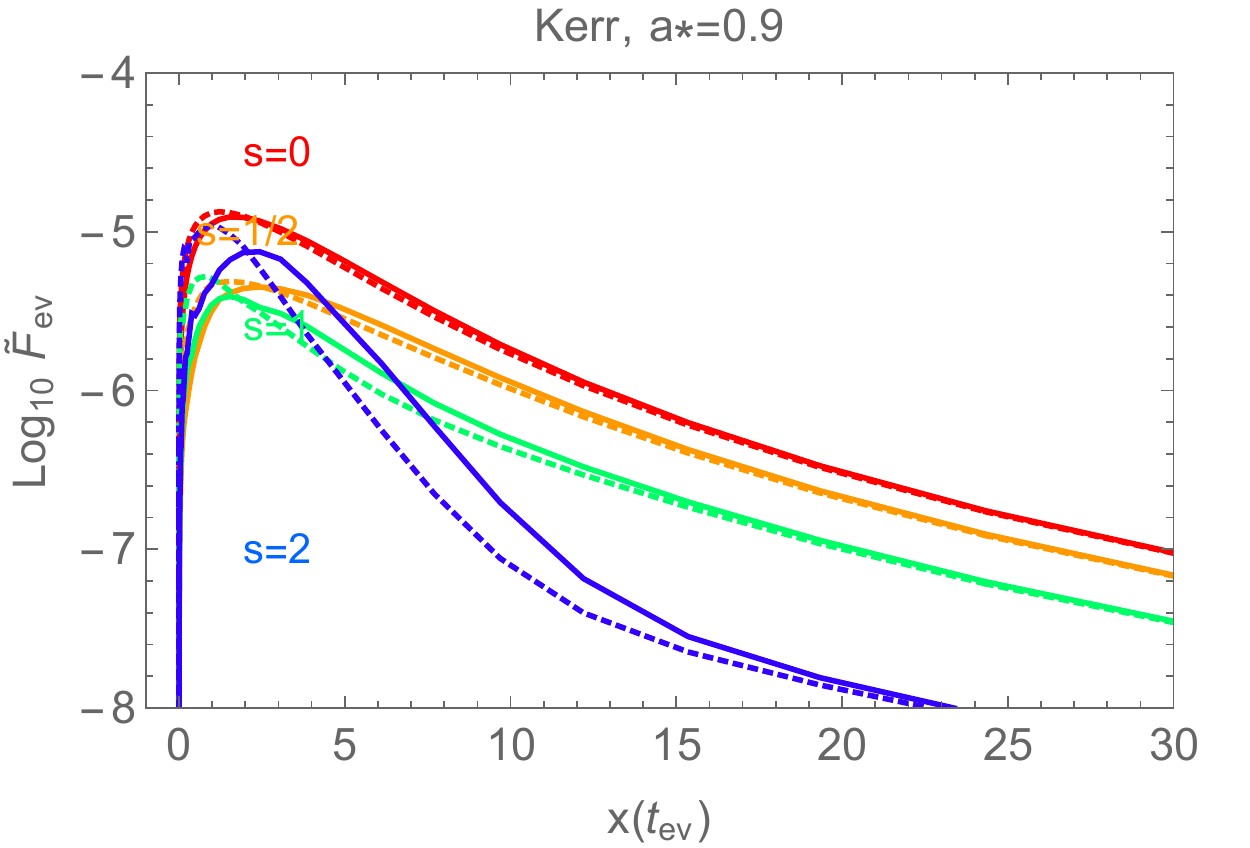}   
\end{center}
\caption{\baselineskip=15 pt \small  
Adimensional function $\tilde F(x(t_{ev)})$, as defined by eq.\,(\ref{eq-tFev}), assuming radiation domination with $\beta=\bar \beta$ (solid) or full BH domination (dashed), calculated using \texttt{BlackHawk}\,\cite{Arbey:2019mbc} within the SM and for various particle spins, as indicated. 
Top left: Schwarzschild case. Top right and bottom: Kerr case with $a_*=0.5$, $a_*=0.8$ and $a_*=0.9$.}
\label{fig-distr-ev}
\vskip .2 cm
\end{figure}

For the Schwarzschild case, previously studied in ref.\,\cite{Auffinger:2020afu}, 
the quantity $\tilde F_{s_i}(x(t_{ev}))$ derived from \texttt{BlackHawk}\,\cite{Arbey:2019mbc} is shown in the top left panel of fig.\,\ref{fig-distr-ev}, 
assuming radiation domination with $\beta = \bar \beta$ (solid) and full (namely a long period of) BH domination (dotted). 
The suppression due to the different values of the spin is manifest. 
The other panels show the same for the Kerr case, taking different initial spins, $a_*=0.5, 0.8, 0.9$, as indicated.

Since the particle momentum scales in time as the inverse of the scale factor, the contribution at small $x(t_{ev})$ is dominated by evaporation in the first period of the BH lifetime, while the contribution at large $x(t_{ev})$ is dominated by evaporation in the last period of the BH lifetime, 
for which the momentum is not much suppressed.
The contribution at small $x(t_{ev})$ is thus particularly sensitive to the Kerr regime: the higher is the value of BH spin parameter $a_*$, 
the more the contribution from particles with high spin, like $s=2$ and $s=1$, gets enhanced. This fact is well known, in particular for gravitons\,\cite{Page:1976ki}. For large $x(t_{ev})$ the four plots of fig.\,\ref{fig-distr-ev} are instead similar, because the Kerr BH has already slowed down its rotation, 
so that the evaporation at late times resembles the Schwarzschild case.

\section{Particle densities at evaporation}
\label{sec-PD}

In this section we calculate the particle number and energy densities for a Kerr BH, which will be useful in the following sections.

\subsection{Number densities at evaporation}

The density at the evaporation time of the particles of type $i$ emitted in the evaporation of the BHs, is given by
\beq
n_i (t_{ev}) =\int dE \,  \, \frac{dn_i}{dE} (t_{ev}) ={n_{BH}(t_{ev})} \, \int dE\, \, \frac{dN_i}{dE}(t_{ev}) =
 n_{BH}(t_{ev}) \,N_i \,\, ,
\eeq
where 
$N_i$ is the total number of $i$ particles produced in the evaporation of a single BH.

\begin{figure}[h!]
\vskip .2 cm 
 \begin{center}
\includegraphics[width= 7.6 cm]{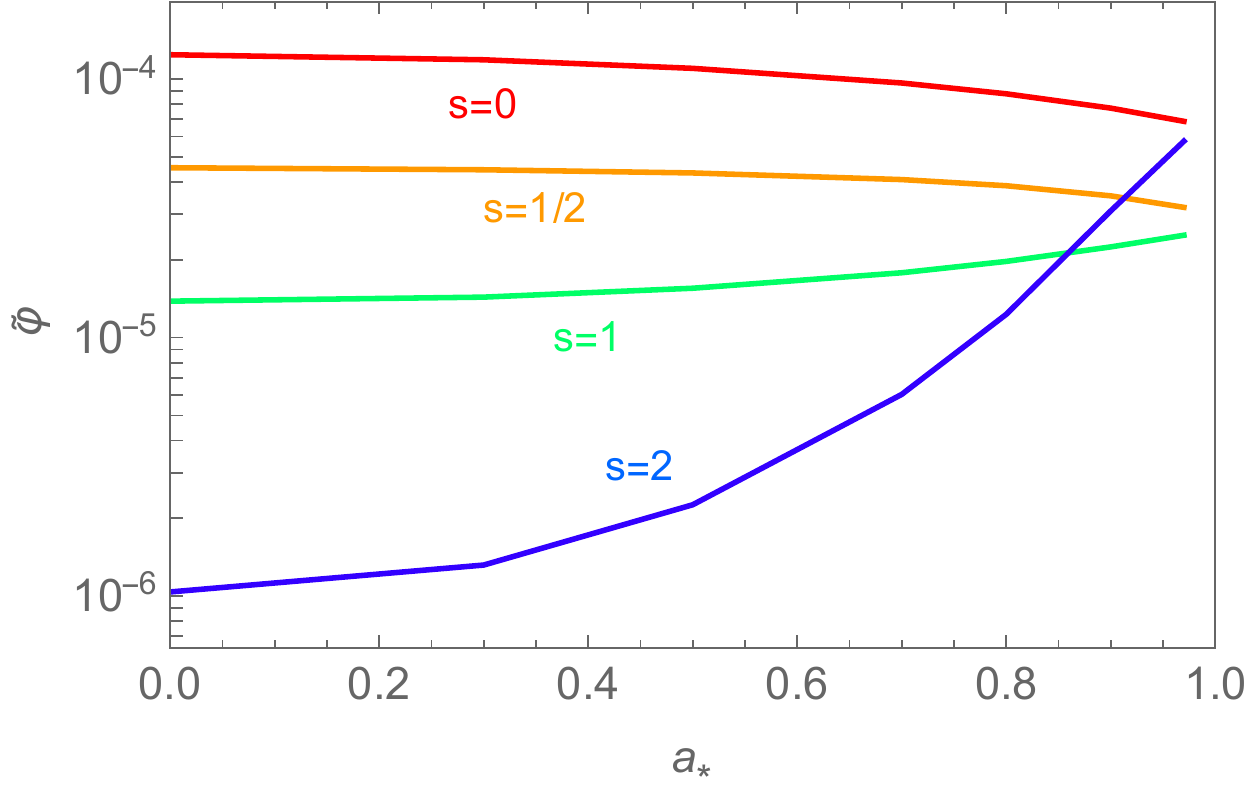}   \,\,\includegraphics[width= 7.6 cm]{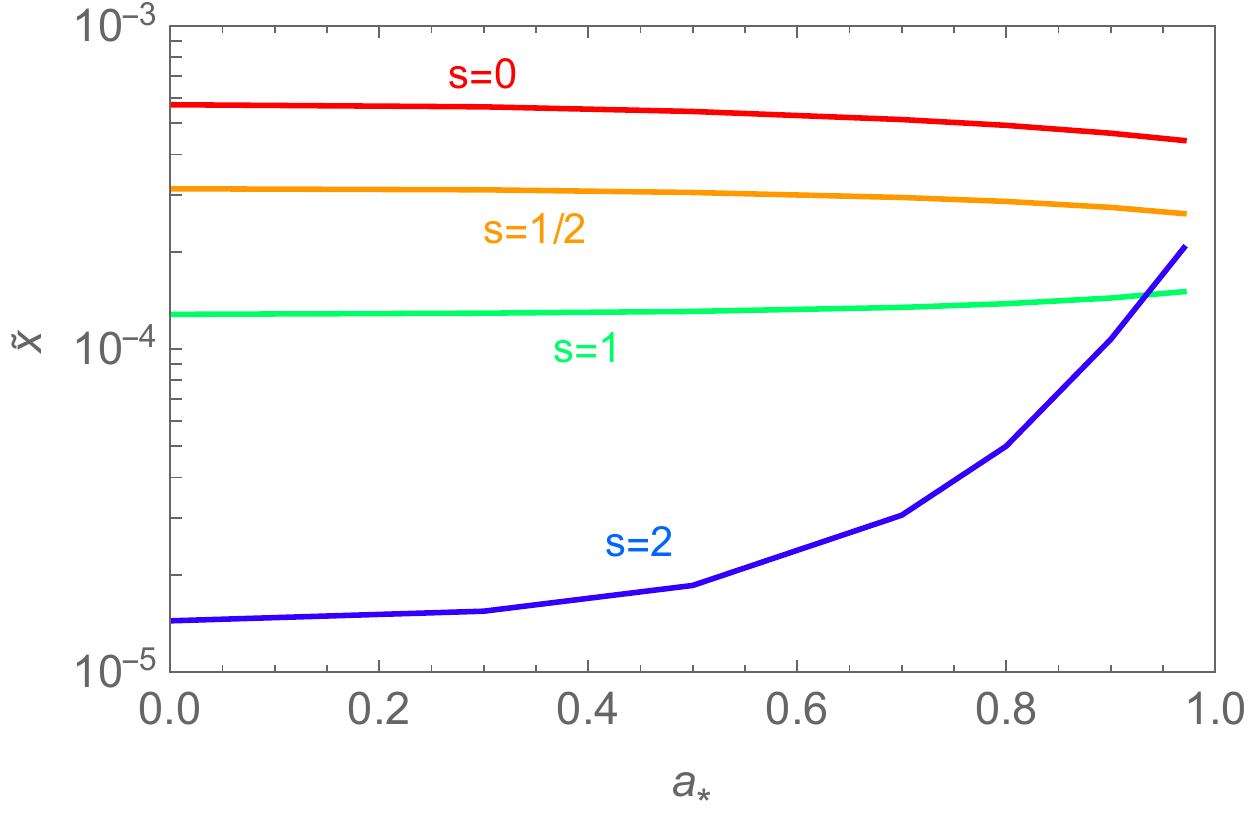}  
\end{center}
\caption{\baselineskip=15 pt \small  
Dependence on $a_*$ of the quantities $\tilde \phi_{s_i}$ and $\tilde x_{s_i}$, for various particles spins, as indicated. 
Radiation domination and BH domination give nearly indistinguishable results.}
\label{fig-tphi}
\vskip .2 cm
\end{figure}

For the relativistic regime, the calculation of $\frac{dN_i}{d(cp)}(t_{ev})$ was done in the previous section, eq.\,(\ref{eq-Fev}).
Using also eq.\,(\ref{eq-tFev}), we have
\beq
N_i 
=(8 \pi)^2 \frac{M_{BH}^2}{M_{Pl}^2} \,  g_i \,  {\tilde \phi_{s_i}} \,\,\, , \,\,\,\,\,\,\,\, {\tilde \phi_{s_i}}=\int_0^\infty dx(t_{ev})\,  \tilde F_{s_i}(x(t_{ev})) \,\, .
\label{eq-NXphi}
\eeq
We show in the left panel of fig.\,\ref{fig-tphi} the quantity $\tilde \phi_{s_i}$: notice that it does not depend on the BH mass, 
but only on the particle spin $s_i$ and the initial spin of the BH, $a_*$.  
It turns out that $\tilde \phi_{s_i}$ is not appreciably different for radiation or BH domination (the difference being at the percent level).
 We can see the remarkable increase in $\tilde \phi_{s_i}$ going from Schwarzschild to extremal Kerr BHs  for $s_i=2$; 
 for lower particle spins, the dependence on $a_*$ is instead mild.

It is interesting to compare the results of the left panel of fig.\,\ref{fig-tphi} for $a_*=0$, with an approximation, obtained in the Schwarzschild case by adopting the geometrical optics approximation (that differentiates only bosons and fermions). 
Neglecting redshift effects and using the geometrical optics approximation, we have
\beq
 N_i =\int_{t_f}^{t_{ev}} dt\,  \int_0^\infty  dE\,   \frac{d^2N_i}{dt dE} \,\, 
 = \frac{1}{2 f_S(M_{BH})}  \frac{27 \zeta(3)}{8^3 \pi^4 }  \,\{1,\frac{3}{4}\} \,g_{i}    \frac{M_{BH}^2 }{M_{Pl}^2 } 
 \,\, ,
\label{eq-GOsf}
\eeq
where the first (second) factor in the curly brackets stands for bosons (fermions).
In particular, comparing with eq. (\ref{eq-NXphi}), we identify
\beq
\,  \,  {\tilde \phi_{s_i}} \leftrightarrow \frac{1}{(8 \pi)^2}   \frac{1}{2 f_S(M_{BH})}  \frac{27 \zeta(3)}{8^3 \pi^4 }  \, \,\{1,\frac{3}{4}\}  
= \{1.2 , 0.90 \} \,\times \,10^{-4}\,\,,
\eeq
which is quite good in the bosonic case, but not very satisfctory in the fermionic one.

\subsection{Energy densities at evaporation}

The energy density of the particle $i$ at evaporation is
\beq
\rho_i (t_{ev}) =\int dE \,\frac{E}{c^2}  \, \frac{dn_i}{dE} (t_{ev}) =\frac{n_{BH}(t_{ev})}{c^2} \, \int dE\, E\, \frac{dN_i}{dE}(t_{ev}) 
= \frac{n_{BH}(t_{ev})}{c^2} \,N_i <E_i(t_{ev})> \,\, ,
\label{eq-rhoXev}
\eeq
where $<E_i(t_{ev})>$ is the mean energy of the $i$ particle at evaporation.

In the relativistic regime, we use eqs. (\ref{eq-Fev}) and (\ref{eq-tFev}), to obtain
\beq
N_i <E_i(t_{ev})>
= (8 \pi) (M_{BH} c^2)  \, g_i \,  {\tilde x_{s_i}} \,\, , \,\,\,\,\,\,\,  \tilde x_{s_i}= \int_0^\infty dx(t_{ev})\, x(t_{ev})\,  \tilde F_{s_i}(x(t_{ev})) \,\, .
\label{eq-tx}
\eeq
As it was the case for $\tilde \phi_{s_i}$, also $\tilde x_{s_i}$ depends on $a_*$ and the particle spin, but not on the BH mass.

The right panel of fig.\,\ref{fig-tphi} shows the numerical values of $\tilde x_{s_i}$ and its dependence on $a_*$.
The results are obtained for radiation domination, but $\tilde x_{s_i}$ is only marginally different for BH domination. 
We can see a remarkable increase for extremal BHs for $s_i=2$ particles, while for lower particle spins the dependence on $a_*$ is mild.

It is interesting to compare the results of the plot for $a_*=0$, with an approximation obtained in the Schwarzschild case and neglecting the redshift effect, 
so that, instead of eq.\,(\ref{eq-Fev}), one uses
\beq
 \frac{dN_i}{d(cp)}(t_{ev})
=\int_{t_{em}}^{t_{ev}} dt \,  \frac{d^2 N_i}{ dt\,d(c p(t)) }  \left( {cp(t)}, T_{BH}(t) \right) \,\,   ,
\eeq
and the energy density of eq.\,(\ref{eq-rhoXev}) becomes instead
\beq
\rho_i (t_{ev})  
=\frac{n_{BH}(t_{ev})}{c^2} \,  \frac{c^4 M_{Pl}^4}{\hbar}\,\int_{t_{em}}^{t_{ev}} dt \,   \frac{g_i\, f_i(M_{BH}(t))}{M_{BH}(t)^2} 
 =n_{BH}(t_{ev})  \, M_{BH}\, \frac{g_i \,f_{i,S}(M_{BH})}{f_S(M_{BH})}  \,\, .
\eeq
where $g_i f_{i,S}(M_{BH})$ is the contribution to the Page function $f_S(M_{BH})$ by the $i$ particle, which in the last equality was
assumed to be constant over the BH lifetime.
Recalling the definition of $\tilde x_{s_i}$, eq. (\ref{eq-tx}),
we can identify 
\beq
  \, \tilde x_{s_i}  \leftrightarrow \frac{1}{ 8 \pi} \,\frac{f_{i,S}(M_{BH})}{f_S(M_{BH})} = \{ 7.0,3.9,1.6,0.18\}\times\,  10^{-4}\,\,\,,\,\, {\rm for }\,s=0,1/2,1,2\,\, ,
\eeq
where the numerical values in the right hand side have been obtained within the SM (that is for $M_{BH}<10^{10}$\, g). 
Notice the quite good agreement with the numerical values of fig.\,\ref{fig-tphi} for $a_*=0$: this demonstrates that the inclusion of the redshift effect is not dramatic.

\begin{figure}[h!]
\vskip .5 cm 
 \begin{center}
 \includegraphics[width= 7.6 cm]{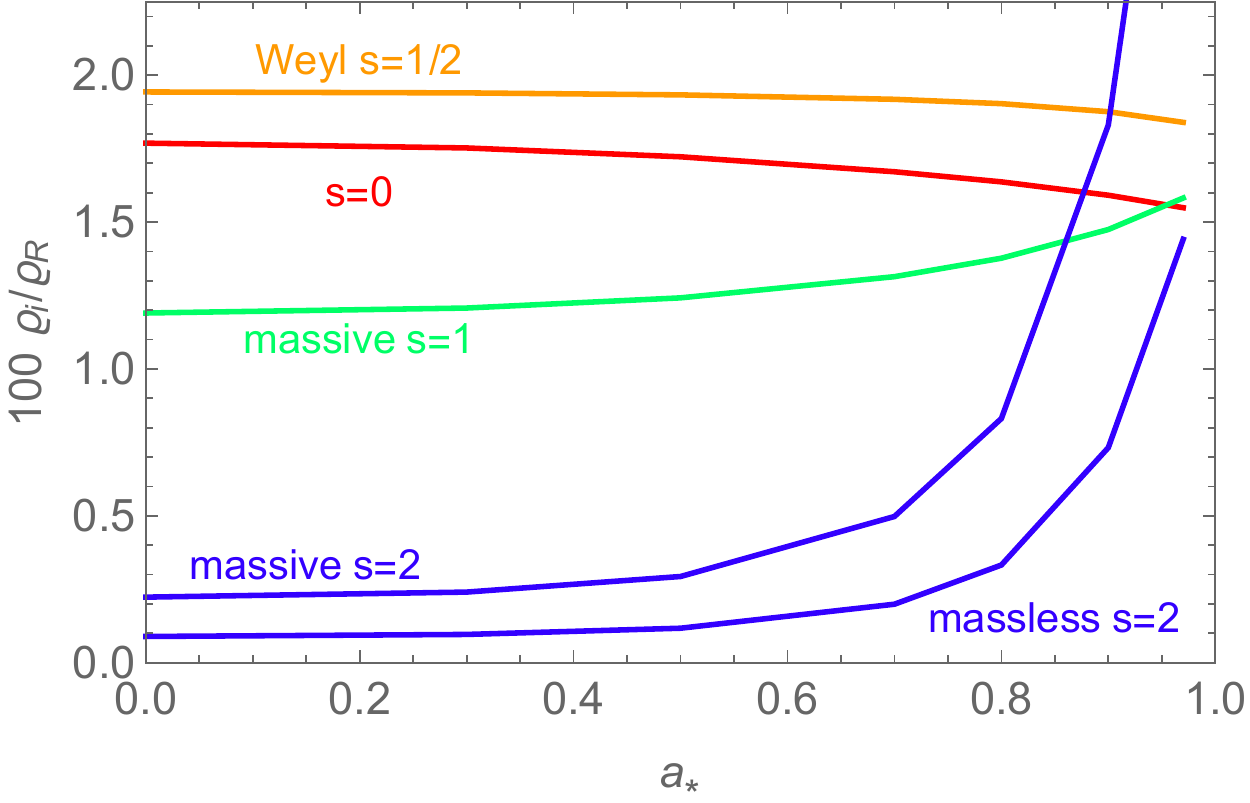}   \,\,\,\,\,\,  \,\,\includegraphics[width= 7.6 cm]{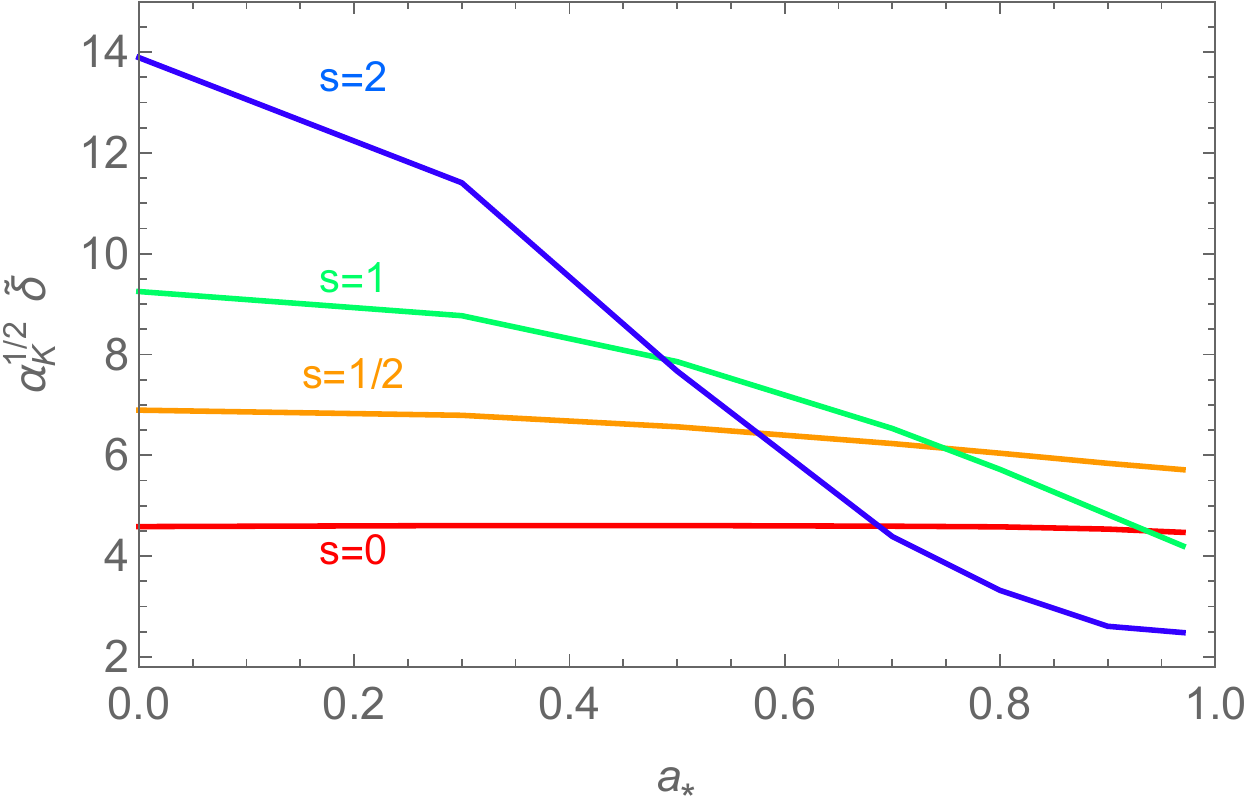} 
\end{center}
\caption{\baselineskip=15 pt \small  
Left: The fraction of the energy density of the $i$ species with respect to the energy density of radiation (that is all the SM particles) at the evaporation,  
${\rho_i(t_{ev})}/ {\rho_{R}(t_{ev})}$, as a function of $a_*$ and for BH domination. We consider a boson ($g_i=1$), a Weyl fermion ($g_i=2)$,
a massive vector ($g_i=3$), a massive ($g_i=5$) and a massless ($g_i=2$), graviton.
Right: The dependence of $\alpha_K^{1/2} \tilde \delta_{s_X}$ on the BH spin $a_*$, for various particle spins, as indicated.}
\label{fig-td}
\vskip .2 cm
\end{figure}

In the case of BH domination, from eqs.\,(\ref{eq-rhoXev}) and (\ref{eq-tx}), the fraction of the energy density of the $i$ species with respect to the energy density of radiation (that is all the SM particles) at the evaporation is
\beq
\frac{\rho_i(t_{ev})} {\rho_{BH}(t_{ev})} \approx \frac{\rho_i(t_{ev})} {\rho_{R}(t_{ev})}
= \frac{N_{i} <E_{i}(t_{ev})>}{\sum_{j=R} N_j <E_j(t_{ev})>} 
=\frac{g_{i}\, \tilde x_{s_{i}}}{ \sum_{j=R}  g_{j} \,\tilde x_{s_j} } \,\, ,
\label{eq-rhoiR}
\eeq 
where the sum over radiation includes all the SM. This is shown the left plot of fig.\,\ref{fig-td}, considering a boson ($g_i=1$), a Weyl fermion ($g_i=2)$,
a massive vector ($g_i=3$), a massive ($g_i=5$) and a massless ($g_i=2$), graviton.
For instance, for a massless (massive) graviton and $a_*=0.7$, we have $\rho_G/\rho_R= 0.20\% (0.50\%)$;
for an extremal BH with $a_*=0.97$, the latter values increase to $1.4\% (3.6\%)$ respectively.

In the case of radiation domination, exploiting the fact that ${\rho_{BH}(t_{ev})} /{\rho_{R}(t_{ev})}=  f(t_{ev})  ={\beta}/{\bar \beta}$, one has 
\beq
\frac{\rho_i(t_{ev})} {\rho_{R}(t_{ev})} 
= \frac{\rho_i(t_{ev})} {\rho_{BH}(t_{ev})}  \frac{\rho_{BH}(t_{ev})} {\rho_{R}(t_{ev})} 
=   \frac{\beta}{\bar \beta}  \frac{g_{i}\, \tilde x_{s_{i}}}{ \sum_{j=R}  g_{j} \,\tilde x_{s_j} } \,\, ,
\label{eq-rhoiBH}
\eeq
namely a global suppression by a factor ${\beta}/{\bar \beta}$ with respect to the results in the left panel of fig.\,\ref{fig-td}.

\section{Viable masses for DR}

In this paper we consider the possibility that, in top of the SM, a particle of the species $X$, with mass $m_X$, is produced in the evaporation of the BH.
If such particle is stable, it would contribute to DM and, if sufficiently light, it would give a significant contribution to DR \cite{Lennon:2017tqq,Hooper:2019gtx,Masina:2020xhk}. Here we generalize to the Kerr case the argument reviewd in ref.\,\cite{Masina:2020xhk} for the Schwarzschild case.

In order for the $X$ particles to give a sizable contribution to DR, their average kinetic energy evaluated at the time of matter radiation equality,
$t_{EQ}$, must exceed their mass: $c p_{EQ} \approx \langle E_X(t_{EQ}) \rangle \gtrsim m_X c^2$. 
Using eqs.\,(\ref{eq-NXphi}) and (\ref{eq-tx}), the average kinetic energy of the emitted $X$ particles is\footnote{It is 
interesting to compare our numerical result with a previous analytical approximation, valid in the Schwarzschild case\,\cite{Fujita:2014hha},
$c p_{ev}  \approx \langle E_X (t_{ev}) \rangle \approx 6 \, k_B T^S_{BH}$.}
\beq
c p_{ev}  \approx <E_X(t_{ev})> 
 =\frac{ \, \tilde x_{s_X} } {\tilde \phi_{s_X}}  \, k_B T^S_{BH} \equiv {\tilde \delta_{s_X}} \, k_B T^S_{BH} \,\, .
 \label{eq-meanE}
\eeq

Since the momentum scales as the scale factor, 
\beq
\langle E_X (t_{EQ}) \rangle 
\approx \langle E_X (t_{ev}) \rangle \, \frac{a(t_{ev})}{a(t_{EQ})}
= \tilde  \delta_{s_X} \, (k_B T^S_{BH})\frac{1}{\alpha'} \frac{k_B T_R(t_{EQ})}{k_B T_R(t_{ev})} \left( \frac{g_{*,S}(t_{EQ})}{g_{*,S}(t_{ev})} \right)^{1/3} \,\, , 
\label{eq-ETEQ}
\eeq
where in the last equality we assumed entropy conservation from evaporation to matter-radiation equality, $\alpha' (s a^3)_{ev}=(s a^3)_{EQ}$.
Using eq.\,(\ref{eq-TevR}), we find the dependence on the BH mass
\beq
 \frac{k_B T^S_{BH}}{k_B T_R(t_{ev})} ={\alpha_K^{1/2}} \, \frac{1}{8 \pi}  \, \left(   \frac{1}{3 f_S(M_{BH})} \right)^{1/2}  
 \left( \frac{16 \pi^3 g_{*}(t_{ev})}{ 45  }     \right)^{1/4}   \left( \frac{     M_{BH} }{  M_{Pl}  }  \right)^{1/2}   \,\, ,
\eeq
so that, assuming $g_{*,S}(t_{EQ}) \approx 3.94$, $ g_{*,S}(t_{ev})\approx 108.75$, taking $k_B T_R(t_{EQ}) \approx 0.75$ eV and $\alpha'=1$, 
the condition to contribute significantly to DR becomes
\beq
m_X c^2 \lesssim \langle E_X (t_{EQ}) \rangle 
\approx 
{\alpha_K^{1/2}}\,  { \tilde \delta_{s_X}} \, \left( \frac{M_{BH}}{1 \,{\rm g}} \right)^{1/2}\,0.11 \,{\rm keV}\, .
\label{eq-EmEQ}
\eeq

The right panel of fig.\,\ref{fig-td} shows the dependence of the quantity $\alpha_K^{1/2} \tilde \delta_{s_X}$ on the BH spin $a_*$. 
For a boson in the Schwarzschild case $\tilde \delta_0\approx 4.5$, so that $m_X c^2 \lesssim 0.50$ keV for $M_{BH}=1$ g, in agreement with ref.\,\cite{Masina:2020xhk}.
For the Kerr case, the suppression induced by factor $\alpha_K^{1/2}$ is mild (at most $0.7$ for extremal BH). Including also the factor $\delta_{s_X}$, we can see from the right panel of fig.\,\ref{fig-td} no significant difference with respect to the Schwarzschild for $s=0,1/2$, while the decrease is significant for $s=1$ and especially for $s=2$.

\section{Stable particles as dark matter}

In this section we assume that the $X$ particle is going to provide the full contribution to the DM observed today.

In the Schwarzschild case, it is well known that there are two possible solutions for DM, denoted as "light" and "heavy" DM \cite{Fujita:2014hha}, 
according to the fact
that the particles are produced during all the BH lifetime or just in its final stages (see e.g. \cite{Masina:2020xhk} and references therein). 
If $m_X c^2 < k_B T^S_{BH}$, the DM candidate belongs to the "light" category, otherwise to the "heavy" one.
In particular, the light DM case is subject to strong constraints from structure formation \cite{Fujita:2014hha, Lennon:2017tqq, Masina:2020xhk, Baldes:2020nuv, Auffinger:2020afu}, so that the BH domination scenario is ruled out.
One might guess \cite{Auffinger:2020afu} that the Kerr case would manage to escape such constraints. 
This what we study in this section, by generalizing the study of the light DM scenario to Kerr BHs.

The cosmological abundance related to the species $X$ at the present time, $t_0$, is proportional to the present number to entropy density 
of such species
\beq
\Omega_{X} =\frac{\rho_{X}   }{\rho_{c}}=\frac{m_{X}  }{\rho_{c}} \frac{n_{X}(t_{0}) }{s(t_{0}) }\, s(t_{0})
=\frac{m_{X} \,s(t_{0})  }{\rho_{c}}   Y_X(t_{0})   \, \,,
\label{eq-OX} 
\eeq
where, defining $H= 100\, h \,{\rm km\,s^{-1}\, Mpc^{-1}}$,
\beq
\rho_c =\frac{3 H^2}{8 \pi G} =1.88 \times 10^{-26} \,h^2 \,{\rm \frac{kg}{m^3}} \,.
\eeq
The entropy density now is obtained from eq.\,(\ref{eq-s}) by putting the CMB temperature $T_{CMB}=2.7255$ K \cite{Akrami:2018odb}:
$s(t_{0})= 2891 /{\rm cm^3}$.
Observationally, the cosmological abundance of cold DM has to be $\Omega_c\approx 0.25$.

One can treat evaporation \cite{Baumann:2007yr} as if all particles were produced at a single instant, $t\approx t_{ev}$. 
The present number-to-entropy density of a stable particle $X$ produced by evaporation is directly related to the BH abundance at evaporation 
 \cite{Baumann:2007yr, Fujita:2014hha} by
\beq
Y_X(t_{0}) =\frac{n_{X}(t_{0}) }{s(t_{0}) }= \frac{1}{\alpha}\frac{n_{X}(t_{ev}) }{s(t_{ev}) }
= \frac{1}{\alpha} N_{X} \frac{n_{BH}(t_{ev})}{s(t_{ev})} = \frac{1}{\alpha} N_{X}  Y_{BH}(t_{ev})\,,
\label{eq-YXtnow}
\eeq
where $\alpha$ parametrizes a possible entropy production after evanescence, $\alpha (s a^3)_{ev}=(s a^3)_{0}$, 
and $N_{X}$ is the number of $X$ particles produced in the evaporation of a single BH\footnote{It is reasonable to assume that entropy is conserved from matter-radiation equality to the present time, so that\,$\alpha \approx \alpha'$.}.
The detailed calculation of $N_X$ was carried out in sec.\,\ref{sec-PD}, while $Y_{BH}(t_{ev})$ can be read from eqs.\,(\ref{eq-YBHR}) and (\ref{eq-YBHBH}). 

\begin{figure}[h!]
\vskip .0 cm 
 \begin{center}
 \includegraphics[width= 7.6 cm]{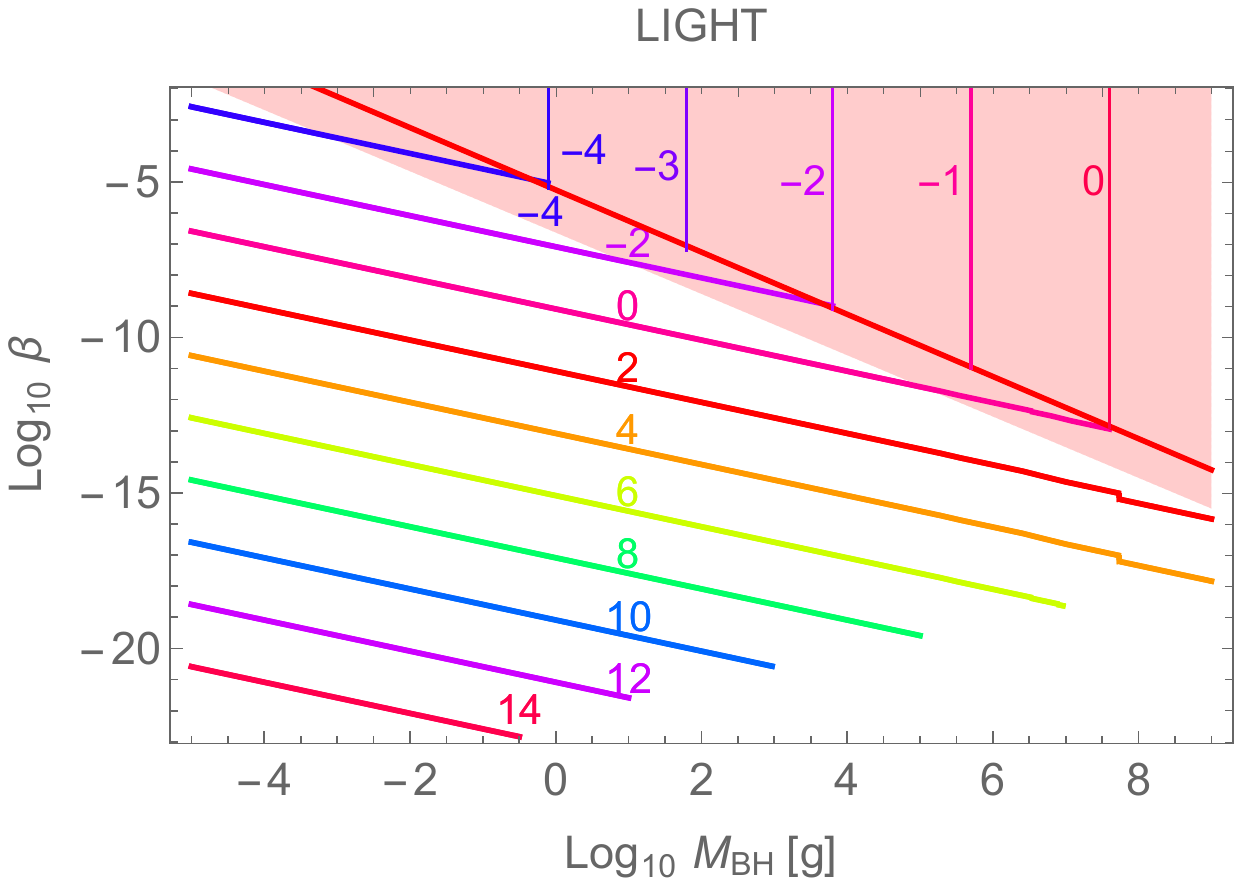}    \,\, 
\includegraphics[width= 7.6 cm]{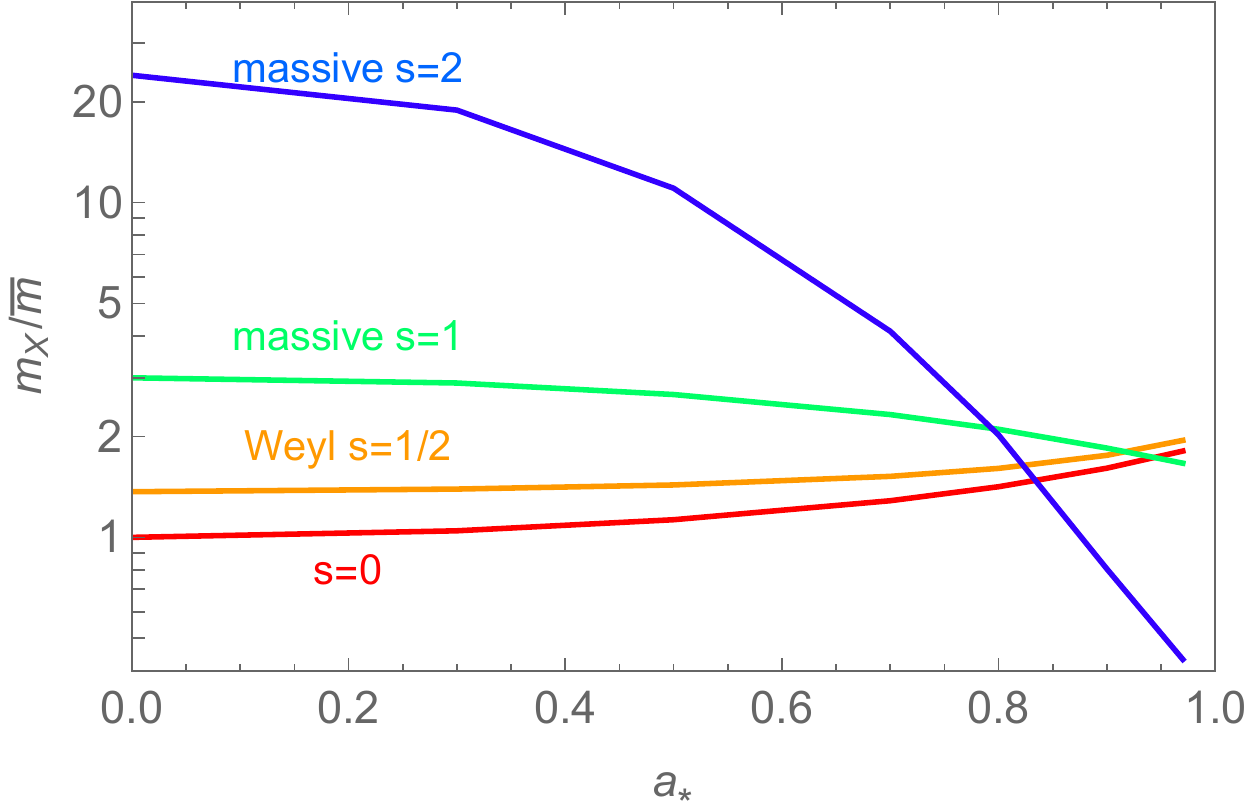}   
\end{center}
\caption{\baselineskip=15 pt \small  Left: Isocontorurs of ${\rm Log}_{10} \bar m c^2$[GeV] for a boson with $g_X=1$, in the Schwarzschild case, 
from ref.\,\cite{Masina:2020xhk}. The shaded region, covering the region of BH domination and a small portion of radiation domination, is ruled out by constraints on structure formation.  
Right: Ratio $m_X/\bar m$ as a function of $a_*$, for radiation domination, with $g_X=1,2,3,5$ and $s_X=0,1/2,1,2$ respectively. }
\label{fig-DM}
\vskip .2 cm
\end{figure}

As reference DM case, we consider a scalar boson (with $g_X=1$) from a Schwarzschild BH. 
The associated value of the DM mass giving the full contribution to DM is denoted by $\bar m$.
We display the iso-contours of ${\rm Log}_{10} \bar m c^2$[GeV]  in the left panel of fig.\,\ref{fig-DM}, taken from \cite{Masina:2020xhk}.
The shaded area, which fully includes the region of BH domination ($\beta>\bar \beta$), is excluded by the constraints on structure formation, as we are going to discuss in the following.

Let us call $m_X$ the mass in the case of a Kerr BH with spin $a_*$, emitting the $X$ particle of spin $s_X=0,1/2,1,2$ (with $g_X=1,2,3,5$, respectively),
providing the full contribution to DM.
Using eq. (\ref{eq-NXphi}), the ratio $m_X/\bar m$ turns out to be, for radiation domination, in which case $Y_{BH}(t_{ev})$ is the same in the Kerr and Schwarzschild case,
\beq
\frac{m_X}{\bar m} 
=\frac{N_0}{N_X}
= \frac{\tilde \phi_0(0)}{g_X \tilde \phi_{s_X}(a_*)} \,\, .
\label{eq-mXmbR}
\eeq
In the right panel of fig.\,\ref{fig-DM} we show the ratio $m_X/\bar m$ as a function of $a_*$. Notice that the Schwarzschild case for the various particle spins agrees with the findings of \cite{Auffinger:2020afu}.
In the more general Kerr case, we can see that the ratio $m_X/\bar m$ increases marginally as a function of $a_*$ for $s=0,1/2$, it slightly decreases for $s=1$, while drastically decreases for $s=2$.

For BH domination, using eq. (\ref{eq-YBHBH}) to take into account the difference in $Y_{BH}(t_{ev})$ for the Kerr and Schwarzschild case, one has 
\beq
\frac{m_X}{\bar m} 
= \alpha_K^{1/2} \frac{\tilde \phi_0(0)}{g_X \tilde \phi_{s_X}(a_*)} \,\, .
\label{eq-mXmbBH}
\eeq

\subsection{Constraints on warm DM}
 \label{sec-wdm}
 
If the $X$ particle is going to provide the full contribution to DM, one has to check that it was cold enough not to waste structure formation. 
The $X$ particles are emitted with a distribution of momenta. An argument based on mean quantities \cite{Fujita:2014hha} 
allows to derive a good estimate for the lower value of $m_X$ that would be compatible with structure formation.
We now generalize the argument of \cite{Fujita:2014hha, Masina:2020xhk, Baldes:2020nuv} to the Kerr case, 
in order to inspect if the tension with structure formation is alleviated\footnote{A more sophisticated analysis 
should follow the lines of \cite{Auffinger:2020afu}.}.

The momentum of the $X$ particle is red-shifted by the expansion of the Universe,
\beq
 p_{0} = \frac{a(t_{ev})}{a(t_{0})} p_{ev} = a(t_{EQ})  \frac{a(t_{ev})}{a(t_{EQ})} \frac{\langle E_X (t_{ev}) \rangle}{c}
 =  a(t_{EQ})  \frac{\langle E_X (t_{EQ}) \rangle}{c}  \,\, ,
\eeq
where we used eqs.\,(\ref{eq-meanE}), (\ref{eq-ETEQ}), $a(t_{0})=1$. 
Assuming that it is no more relativistic, the velocity of the $X$ particle now is 
\beq
 \frac{v_X}{c}= \frac{p_{0}}{c \,m_X} 
 =  a(t_{EQ})\, \frac{\langle E_X (t_{EQ}) \rangle}{m_X c^2} 
 =  {\alpha_K^{1/2}} \,  \frac { \tilde \delta_{s_X}} { \tilde \delta_{0}}\, \frac{\bar m }{m_X }   \frac{\bar v}{c}
 \label{eq-vX}
\eeq
where we used eqs.\,(\ref{eq-EmEQ}) and defined $\bar v/c$ to be the velocity of the previously introduced reference DM case, 
that is a scalar (with $g_X=1$) in the Schwarzschild case, so that
\beq
\frac{\bar v}{c} =a(t_{EQ})\, \frac{1}{\bar m c^2} \,  \tilde \delta_{0} \left( \frac{M_{BH}}{1 \,{\rm g}} \right)^{1/2}\,0.11 \,{\rm keV}
= 0.9 \times 10^{-6}\, ,
\eeq
where in the last equation we used $a(t_{EQ})=\Omega_R/\Omega_{M} \approx 1.8\times 10^{-4}$.

For radiation domination, using eq.\,(\ref{eq-mXmbR}), eq. (\ref{eq-vX}) becomes
\beq
 \frac{v_X}{c}=g_X\, {\alpha_K^{1/2}}\,   \frac { \tilde x_{s_X}(a_*)} { \tilde x_{0}(0)}\,  \frac{\bar v}{c}\, , 
 \label{eq-vXR}
\eeq
while for BH domination, using eq.\,(\ref{eq-mXmbBH}), it becomes
\beq
 \frac{v_X}{c}=g_X\, \frac { \tilde x_{s_X}(a_*)} { \tilde x_{0}(0)}\,  \frac{\bar v}{c}\, , 
  \label{eq-vXBH}
\eeq


The velocity $v_X$ is subject to an upper bound that can be estimated to be comparable to the upper limit on the velocity of a warm thermal relic DM candidate.
Assume that a warm thermal relic giving the full contribution to DM was relativistic at decoupling. In order not to waste structure formation, the lower bound on its mass 
is $m_{W} c^2 \gtrsim 3.5$ keV (at 2$\sigma$) \cite{Irsic:2017ixq}.
Since both the velocity and particle temperature scale as the inverse of the scale factor, and assuming entropy conservation from decoupling to the present epoch \cite{Bode:2000gq}
\footnote{If the warm particles decouple when relativistic, their momentum distribution function remains constant until gravitational clustering begins.
All particle momenta scale as $a^{-1}$ which we can describe by scaling their temperature $T_W$ accordingly. 
When the particles become non-relativistic we can use $p = m_W v_W$. 
}
\beq
 \frac{v_{W} }{c} 
 \approx \frac{k_B T_W(t_{0})}{m_{W} c^2} 
=\left( \frac{g_{*,S}(t_{r})}{g_{*,S}(t_{dec})}\right)^{1/3} \left(\frac{4}{11}\right)^{1/3} \frac{k_B T_{CMB}}{m_{W} c^2}\lesssim 0.7\times  10^{-8} \,\, ,
\label{eq-limV}
\eeq
where we used the fact that ${g_{*,S}(t_{r})}/{g_{*,S}(t_{dec})} = 11\, {\rm eV} /(m_W c^2)$ (see e.g. the discussion in \cite{Auffinger:2020afu}).

If thew $X$ particles are going to fully contribute to DM, we have to require $v_X\lesssim v_{W}$. Hence there is a tension with structure formation because $\bar v$ turns out to be bigger than $v_W$ by two orders of magnitude (consistently with the findings of \cite{Fujita:2014hha, Masina:2020xhk,Baldes:2020nuv,Auffinger:2020afu}). This rules out all the region of BH domination, and also a small confining portion of radiation domination, as can be seen from the left panel of fig.\,\ref{fig-DM}.

In the Schwarzschild case with non zero spins, the tension is reduced for increasing spin values, because of the suppression in $v_X$ coming from the factor $\tilde x_{s_X}(0)/\tilde x_0(0)$, see eqs.\,(\ref{eq-vXR}), (\ref{eq-vXBH}) and the right panel of fig.\,\ref{fig-tphi}.
But even for $s=2$ (with $g_X=5$) the tension persists because, as can be seen from fig.\,\ref{fig-tphi}, one has $ g_X \tilde x_2(0)/\tilde x_0{(0)} \approx 0.1$: the tension is thus reduced at the level of one order of magnitude, consistently with the findings of \cite{Auffinger:2020afu}.

In the more general Kerr case, we can derive from the right panel of fig.\,\ref{fig-tphi} what is the behavior of the quantity ${ \tilde x_{s_X}(a_*)}/ { \tilde x_{0}(0)}\, $ for increasing values of $a_*$. 
For $s=0,1/2,1$ there is no significant difference with respect to the Schwarzschild case; for $s=2$,
since $\tilde x_{2}(a_*)$ is an increasing function of $a_*$, the tension with structure formation become definitely worse than in the Schwarzschild case. 

This hows that  there is no possibility to save the BH domination scenario with Kerr BHs.

\section{Stable particles as dark radiation}

Let us assume that the $X$ particle is light enough to contribute to DR (in which case the contribution to DM is marginal).
The contribution of such a DR component to the effective number of relativistic dof is parametrized by 
\beq
\Delta N_{eff} = \frac{\rho_{X}(t_{EQ})}{\rho_{R}(t_{EQ})}   \left(  N_\nu + \frac{8}{7} \left(  \frac{11}{4}\right)^{4/3} \right) \,,
\label{eq-defNeff}
\eeq
where $N_\nu=3.045$ \cite{deSalas:2016ztq}, and $t_{EQ}$ is the time of matter-radiation equality.

We now extend to the Kerr case the argument followed in ref.\,\cite{Hooper:2019gtx, Masina:2020xhk} to calculate $\Delta N_{eff}$.
The ratio of the energy density in DR with respect to radiation at matter-radiation equality is
\beq
\frac{\rho_{X}(t_{EQ})}{\rho_{R}(t_{EQ})} 
=\frac{1} {{\alpha'}^{4/3}} \, \frac{\rho_{X}(t_{ev})}{\rho_{R}(t_{ev})}\,  \frac{   g_{*,S} (t_{EQ}) } {g_*(t_{EQ})   }  \frac{   g_{*,S} (t_{EQ})^{1/3} } {  g_{*,S} (t_{ev}) ^{1/3}  }\,.
\eeq

As shown e.g. in fig.\,3 of \cite{Auffinger:2020afu}, within the SM, $g_{*,S}(t_{ev})=106.75$ is constant for $M_{BH}<10^6$\,g
(while it drops down to the value $10.75$ at $M_{BH}\approx 10^{9}$ g).
Substituting the above expression in eq.\,(\ref{eq-defNeff}), and taking $g_{*,S} (t_{EQ})=3.94$, $g_{*} (t_{EQ})=3.38$, $N_\nu=3.045$, $g_{*,S} (t_{ev})=106.75$, one has 
\beq
\Delta N_{eff} 
\approx 2.89\,\frac{1}{ {\alpha'}^{4/3}} \, \frac{\rho_{X}(t_{ev})}{\rho_{R}(t_{ev})}  \,.
\eeq
In the BH mass range above $10^6$ g, there is a slight enhancement with respect to the factor $2.89$, which has to be substituted by $6.21$ for $M_{BH} = 10^{9}$ g.

The ratio $\rho_{X}(t_{ev})/ \rho_{R}(t_{ev})$ was already studied in eqs. (\ref{eq-rhoiR}), (\ref{eq-rhoiBH}).
For BH domination, within the SM with the additional DR, it was shown in the left panel of fig.\,\ref{fig-td}, for different particle spins. 
For radiation domination the results have to be suppressed by $\beta/\bar\beta$.

\begin{figure}[h!]
\vskip .0 cm 
 \begin{center}
 \includegraphics[width= 7.6 cm]{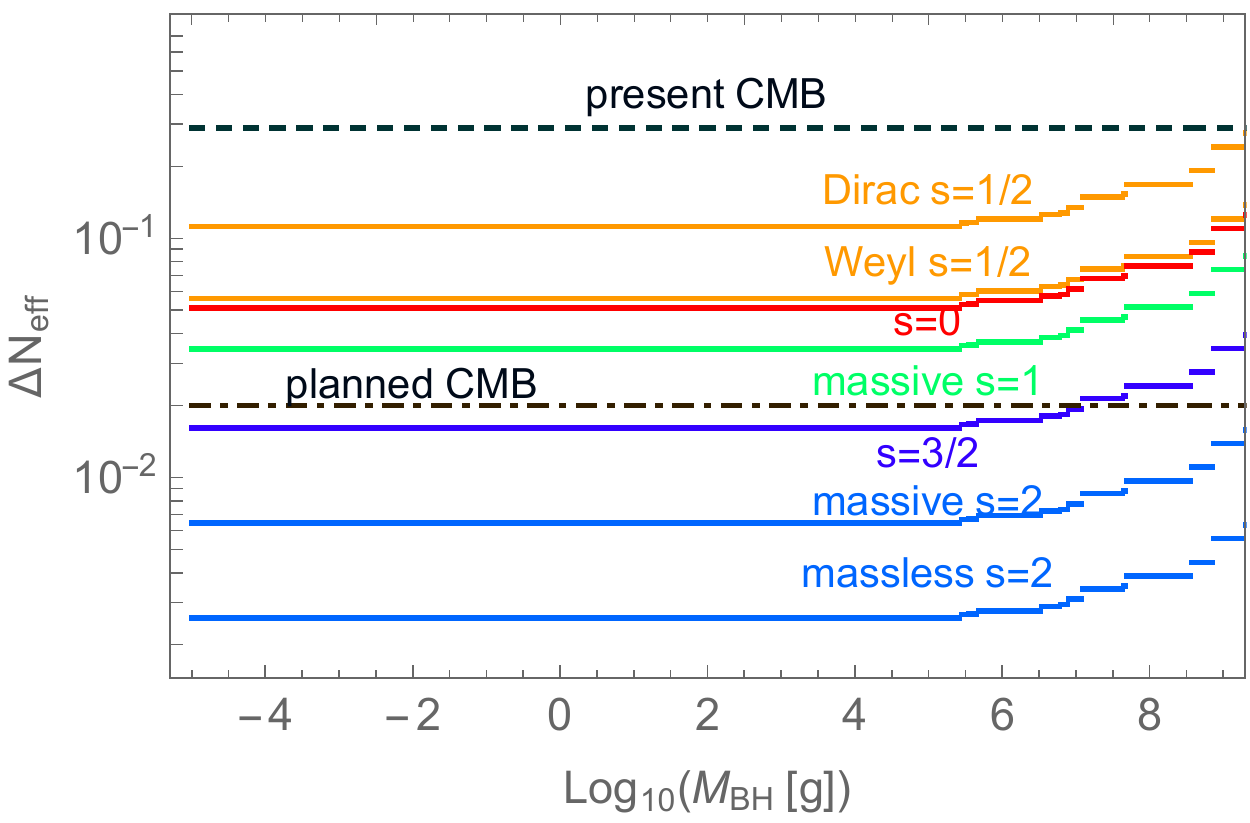}    \,\,\, \includegraphics[width= 7.6 cm]{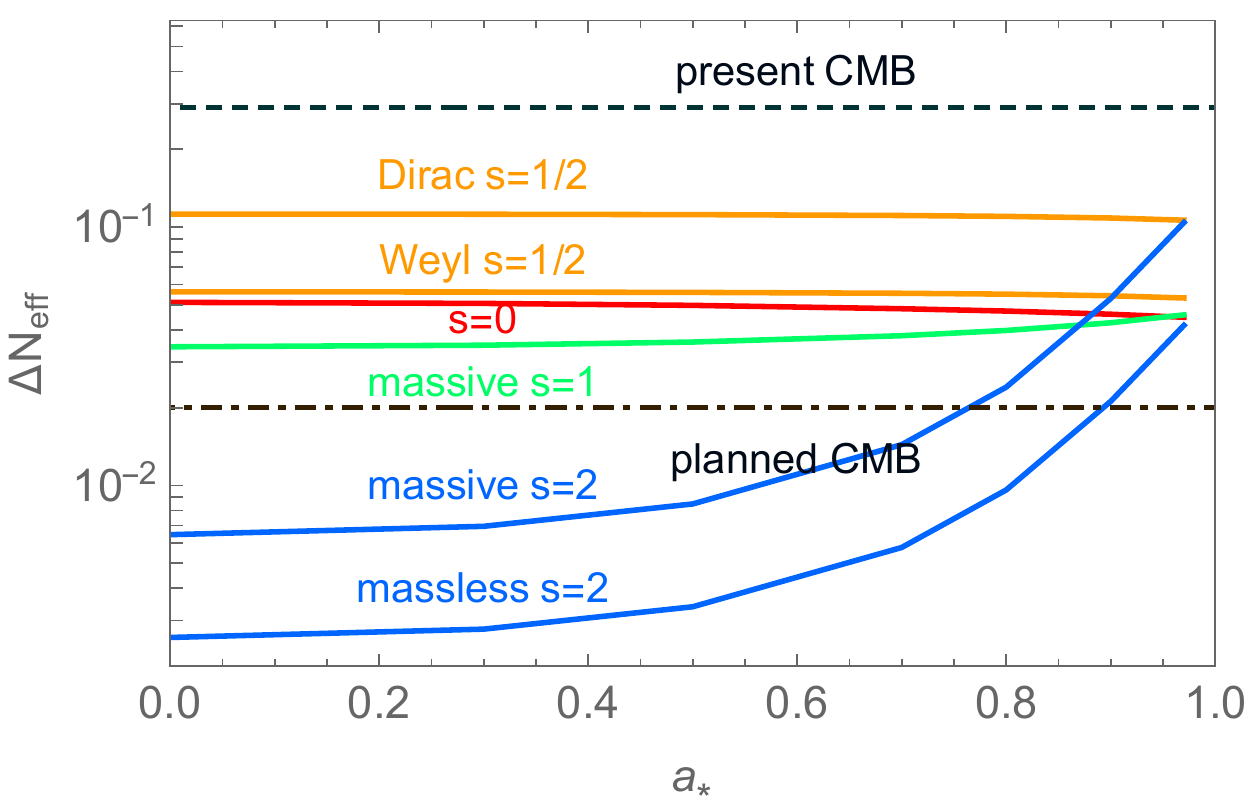}   
\end{center}
\caption{\baselineskip=15 pt \small  
Left: Various DR contributions to $\Delta N_{eff}$, as a function of the BH mass, assuming an epoch of BH domination and taking $\alpha'=1$.
Right: the dependence of  $\Delta N_{eff}$ on the BH spin $a_*$, for $M_{BH}<10^6$\,g.}
\label{fig-DR}
\vskip .2 cm
\end{figure}

For the Schwarzschild case and BH domination, the associated prediction for $\Delta N_{eff}$ is shown in the left panel of fig. \ref{fig-DR}, 
taking $\alpha'=1$, and for various particle spins, as indicated.
These results are in full agreement with \cite{Hooper:2019gtx}.
The present sensitivity to $\Delta N_{eff}$ of CMB observations is shown: since $N_{eff} = 2.99 \pm 0.17$\,\cite{Aghanim:2018eyx},
one has $ N_{eff}<3.33$ at $2\sigma$, or equivalently $\Delta N_{eff} =N_{eff}-N_\nu  <  0.29$ at $2\sigma$.
Interestingly enough, there are optimistic possibilities of detecting some signal in the future \cite{Hooper:2019gtx},
as the predicted contribution to $\Delta N_{eff}$ is potentially within the projected reach of stage IV experiments, 
$\Delta N_{eff} \approx 0.02$: this is the case for DR particles with $s < 3/2$, but not for a massless or massive $s=2$ particle (called graviton for short).

The right panel of fig.\,\ref{fig-DR} shows the dependence of  $\Delta N_{eff}$ on the BH spin, $a_*$.
One can see that for  $s=0,1/2,1$ there are no significant changes, while for $s=2$ there is a significant increase. 
However, for $s=2$, the contribution to $\Delta N_{eff}$ remains below the future observable region, unless 
$a_*>0.9\, (0.75)$ in the massless (massive) case respectively, for $M_{BH}<10^6$\,g. 
In the case of larger values of $M_{BH}$ (corresponding to evaporation just before nucleosynthesis), 
there is a further slight increase, as the left panel of fig.\,\ref{fig-DR} shows.

%

We thus find less optimistic results than those presented in ref.\,\cite{Hooper:2020evu} for "hot" gravitons. 
We comment here on the source of the discrepancy.
The latter work studies the case of BHs that are spinning as a consequence of having undergone
previous mergers. The distribution of angular momenta predicted for such a BH population peaks strongly at $a_*=0.7$,
almost entirely independently of the masses or the initial spin distribution of the merging binaries \cite{Fishbach:2017dwv}.
After integrating over the evolution of BHs with a distribution of initial spins as described in \cite{Fishbach:2017dwv},
ref.\,\cite{Hooper:2020evu} finds that approximately $f_G \approx 0.47\%$ of the energy emitted as Hawking radiation is in the form of massless "hot" gravitons. On the contrary, as appears from the left panel of fig.\,\ref{fig-td}, we find less optimistic values: in particular for a massless (massive) graviton and $a_*=0.7$, we have $f_G=\rho_G/\rho_R= 0.20\% (0.50\%)$; for an extremal BH with $a_*=0.97$, the latter values increase to $1.4\% (3.6\%)$ respectively.
Our significantly smaller estimate for $f_G$ is thus the source of the discrepancy in the associated "hot" graviton contribution to $\Delta N_{eff}$:
indeed, for ref.\,\cite{Hooper:2020evu}, $\Delta N_{eff}$ might be as large as  $0.01-0.03$ for $a_*=0.7$, and even as large as $\Delta N_{eff} =0.3$ for near extremal BHs. According to our findings, this would rather apply to the massive graviton case, but not to the massless one.

\section{Conclusions}

We have extended to the Kerr case the study of DM and DR from evaporating primordial BHs. 

For DM, one might have expected \cite{Auffinger:2020afu} that the constraints from structure formation that exclude the scenario of BH domination for light DM in the Schwarzschild case, would have been softened. We find instead that, while for the lower spins as $s=0,1/2,1$ the tension is not changed significantly for all values of the BH spin $a_*$, for $s=2$ the tension is even enhanced with increasing values of $a_*$. 
We conclude that invoking an angular momentum for the evaporating BHs, does not offer a solution to save the BH domination scenario. 

A couple of possibilities to save BH domination should be mentioned. 
As suggested in \cite{Fujita:2014hha}, some mechanism providing entropy non conservation and taking place after the evaporation of primordial BHs (like \emph{e.g.} moduli decay) might succeed this task. 
In this work we considered non-interacting DM from primordial BHs evaporation, but allowing for self-interacting DM offers the possibility to escape the structure formation bound in the light case for BH domination \cite{Bernal:2020kse}: thermalization in the DM sector decreases the mean DM kinetic energy and, together with number-changing processes, can have a strong impact, in particular enhancing the DM relic abundance by several orders of magnitude.

For DR, it is well known \cite{Hooper:2019gtx, Masina:2020xhk} that in the Schwarzschild case particles with $s=0,1/2,1$ might give a contribution to 
$\Delta N_{eff}$ at hand of future experimental sensitivity, while this does not applies to the massive or massless $s=2$ case (see the left panel of fig.\,\ref{fig-DR}).
In the Kerr case, we find that the contribution to $\Delta N_{eff}$ by particles with spin $s=0,1/2,1$ has a very mild dependence on $a_*$, while for $s=2$ the contribution significantly increases with $a_*$ \,\cite{Hooper:2020evu}. 
In any case, for $s=2$ and a moderate value of the spin parameters like $a_*=0.7$, we find that $\Delta N_{eff}$ remains below the projected sensitivity in the massless case (while in the massive case it would reach the level of the planned sensitivity only for BHs evaporating just before nucleosynthesis),
see the right panel of fig.\,\ref{fig-DR}. In the massless case for $s=2$, only for extreme values of the spin parameter, $a_*>0.9$, $\Delta N_{eff}$ would reach the projected experimental sensitivity. Our results are thus less optimistic than those of ref.\,\cite{Hooper:2020evu} for the "hot" graviton case.

\section*{\Large Acknowledgements}

I.M. acknowledges partial support by the research project TAsP (Theoretical Astroparticle Physics) funded by the Istituto Nazionale di Fisica Nucleare (INFN). 

\bibliographystyle{elsarticle-num} 
\bibliography{bib} 
\end{document}